\documentclass[12pt,a4paper,oldfontcommands]{book}
\usepackage{setspace}
\usepackage[utf8]{inputenc}
\usepackage[T1]{fontenc}
\usepackage{microtype}
\usepackage[dvips]{graphicx}
\usepackage{xcolor}
\usepackage{times}
\usepackage{amsmath,amssymb,amsfonts}
\usepackage{mathtools, nccmath}
\DeclarePairedDelimiter\ceil{\lceil}{\rceil}

\usepackage{bm}   
\usepackage{amsthm}

\newtheorem*{remark}{Remark}
\newtheorem{assumption}{Assumption}

\theoremstyle{definition}
\newtheorem{example}{Example}[section]
\newtheorem{definition}{Definition}[section]
\usepackage[linesnumbered,ruled,longend]{algorithm2e}
\usepackage{placeins}
\usepackage{lscape}
\usepackage{longtable}

\usepackage{etoolbox}
\AtBeginEnvironment{align}{\setcounter{equation}{0}}
\usepackage{chngcntr}
\counterwithout{equation}{chapter} 
\def\code#1{\texttt{#1}}

\usepackage{pythonhighlight}
\usepackage{fancyvrb}
\usepackage{subcaption,siunitx,booktabs}

\usepackage[
breaklinks=true,colorlinks=true,
linkcolor=black,urlcolor=black,citecolor=black,
bookmarks=true,bookmarksopenlevel=2]{hyperref}

\usepackage{fix-cm}
\usepackage{xcolor}
\usepackage{titlesec}
\definecolor{gray75}{gray}{0.75}

\titleformat{\chapter}[display]{\filleft}{\fontseries{b}\fontsize{100}{130}\selectfont\textcolor{gray75}\thechapter}{0pt}{\Huge\bfseries}[]%

\begin{document}
\setcounter{page}{1}
\pagenumbering{roman}
%
%
\thispagestyle{empty}

{
\sffamily
\centering
\Large

\includegraphics[scale=0.8]{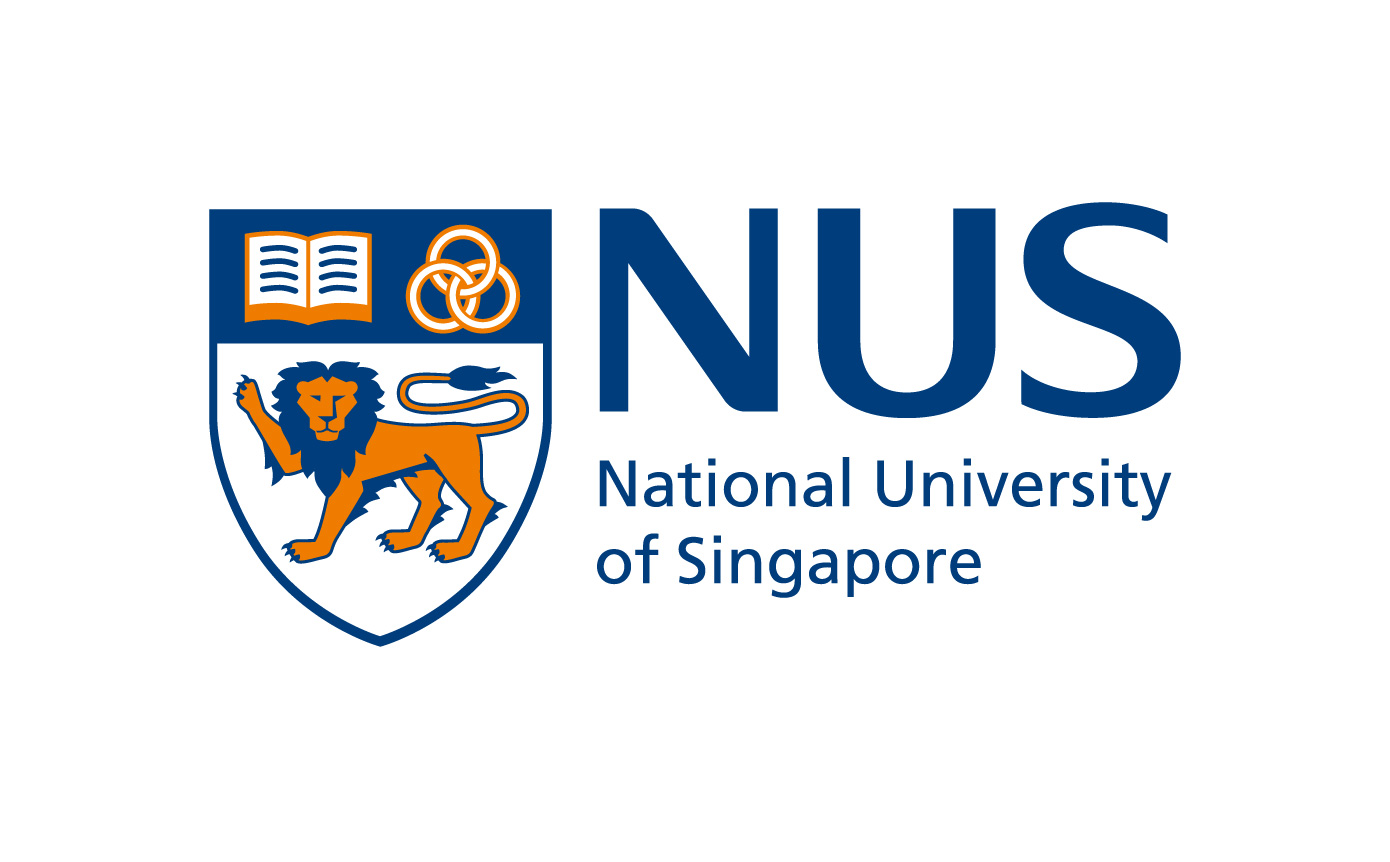}

{\huge 
Solving the Joint Order Batching and Picker Routing Problem for Large Instances
}

\vspace{2.5cm}

{\LARGE
Khoong Wei Hao
}

\vspace{2.5cm}

A thesis submitted in partial fulfillment for the

degree of Bachelor of Science

in Applied Mathematics

\vspace{1em}

National University of Singapore

\vspace{1.22cm}

Supervisors:\vspace{2mm}\\
\resizebox{\textwidth}{!}{Dr Melvin Zhang \& Assistant Professor Pang Chin How Jeffrey}

}

\cleardoublepage

\thispagestyle{empty}
\vspace*{\fill}
\noindent To my family, in particular my father and late mother who made many sacrifices to raise me into the individual I am today, my friends whom showered me with continuous care and support over the years, and also my teachers.
\vspace*{\fill}

\cleardoublepage

\section*{Acknowledgements}

Firstly, I would like to express my sincere gratitude to my supervisor Dr. Zhang, CTO of Cosmiqo International, for the continuous support, his dedication, patience, guidance, and the incredible, meaningful experience I gained working with him on the project. His dedication and guidance helped me in all the time of the project and writing of this thesis. I have gained substantial amounts of advice and insights from him, especially in the art of being an efficient programmer. I certainly do believe and hope that from this point on, I would have gone beyond the stage of being a novice programmer. 

I would also like to thank my co-supervisor Assistant Prof. Pang, who has given me valuable advice throughout the project, and also checked on me during the toughest of times which I have experienced mid-way through the project. In particular, I would like to extend my thanks to Prof. Zhang, head of the honours project committee at the department of Mathematics, for liaising with both Assistant Prof. Pang and Dr. Zhang and presenting me the opportunity to take up the project.

I would also like to express my heartfelt thanks to all my friends who have not only supported me emotionally throughout the project, but also throughout the years of my education. 

Last but not least, I would like to thank my family for supporting me both emotionally and spiritually throughout the write-up of this thesis, and my life as well. 

\cleardoublepage

\Large
 \begin{center}
Solving the Joint Order Batching and Picker Routing Problem for Large Instances\\ 

\hspace{10pt}

\large
Khoong Wei Hao$^1$, Melvin Zhang$^2$\\

\hspace{10pt}

\small  
$^1$Department of Mathematics,\\
National University of Singapore,\\
Level 4, Block S17, 10 Lower Kent Ridge Road,\\
Singapore 119076,\\
khoongweihao@u.nus.edu\\
\vspace{1mm}
$^2$Cosmiqo International Pte Ltd\\
melvin@cosmiqo.com

\end{center}

\hspace{5pt}

\normalsize

\begin{center}
\textbf{Abstract}
\end{center}

In this work, we investigate the problem of order batching and picker routing in warehouse storage areas. These problems are known to be capital and labour intensive, and often contribute to a sizable fraction of warehouse operating costs. Here, we consider the case of online grocery shopping where orders may consist of dozens of items.

We present the problem introduced in~\cite{valle2017} and tackle the issue of solving the problem heuristically with proposed methods of solving that utilize batching and routing heuristics. Instances with up to 50 orders were solved heuristically in large simulated warehouse instances consisting of 8 to 30 aisles, with 1 to 4 blocks. The proposed methods were shown to have relatively short computation times as compared to optimally solving the problem in~\cite{valle2017}. In particular, we showed that a proposed method which utilizes an optimal solver for routing yielded poorer results than methods that utilize routing heuristics.\\

\noindent\textbf{Keywords:} integer programming, inventory management, order batching, order picking, picker routing

\tableofcontents

\listoffigures

\listoftables

\cleardoublepage

\setcounter{page}{1}
\pagenumbering{arabic}

\chapter{Introduction}

\begin{doublespacing}

\section{The Business Problem}
The project undertaken in this thesis investigates methods to optimize order picking by batching several orders together and by planning a good route to minimize the distance required to pick up all the items. The proposed methods scale up to typical warehouse sizes and are implemented and experimentally tested on real data sets. This project is in collaboration with an industry partner, Cosmiqo International Pte Ltd\footnote{See https://cosmiqo.com/}. 

We thus answer the following two main questions in this thesis:

\begin{enumerate}
\item Across all order and warehouse instances, what is the quality of solution vs time trade-off?
\item If I have a warehouse with $X$ aisles and $Y$ cross-aisles and a number of orders to batch, what method should be used to solve the problem within the time limit?
\end{enumerate}

\section{Warehouse Operations}
In a warehouse, reorganization and repackaging of products are performed. A product typically arrives at the warehouse packaged in large quantities and leaves packaged in smaller quantities. This also means that one of the important roles of the warehouse is to receive large quantities of products and to redistribute them in smaller quantities. Here, products can come from manufacturers in pallet-size quantities, but get shipped out to customers in case/pack quantities.

Reorganization of products is carried out via the inbound (receiving, put-away) and outbound (order-picking, checking, packing, shipping) processes in sequence~\cite{bartholdi2017}:
\begin{itemize}
\item \textit{Receiving}: Consists of unloading of products from transportation vehicles from a manufacturer to receiving docks, inspection of products for damages or for missing products, and updating existing warehouse inventory records to reflect changes in stock.

\item \textit{Put-Away}: Involves the movement of products from receiving docks to their allocated storage locations, a shipping dock, other locations in a warehouse, and also the movement of products between these areas.

\item \textit{Order-Picking}: Defined as the process of retrieving products (from orders) from storage blocks in a warehouse in response to specific customer requests~\cite{valle2017, koster2007}.

\item \textit{Checking \& Packing}: Checking is a labor-intensive process that checks whether each customer's order is complete and accurate. Packing, which comes shortly after checking, involves collating items in a customer's order for shipping. Packing is a complicated process as customers generally prefer to receive all items in their order in a minimal number of packages, as this minimizes shipping and handling costs. 

\item \textit{Shipping}: Includes loading of products onto transportation vehicles, the inspection of products to be shipped to customers, and also the updating of warehouse inventory records. 
\end{itemize}

The importance of determining an appropriate storage location  is high as the location of a product largely determines how quickly you retrieve it for a customer, and also the cost of retrieving it at a later time. Thus, a second inventory of storage locations must be managed on top of product inventories. 

\section{Order-Picking}
Order picking is the most challenging of operations in most warehouses, as it is the most labour intensive operation and it also determines the quality of service experienced by customers on the ground. In particular, it typically accounts for as much as $60\%$ of all labor activities~\cite{gademann2005}, and in general order-picking time broken down as such~\cite{bartholdi2017}:

\begin{center}
\includegraphics[width=.65\textwidth]{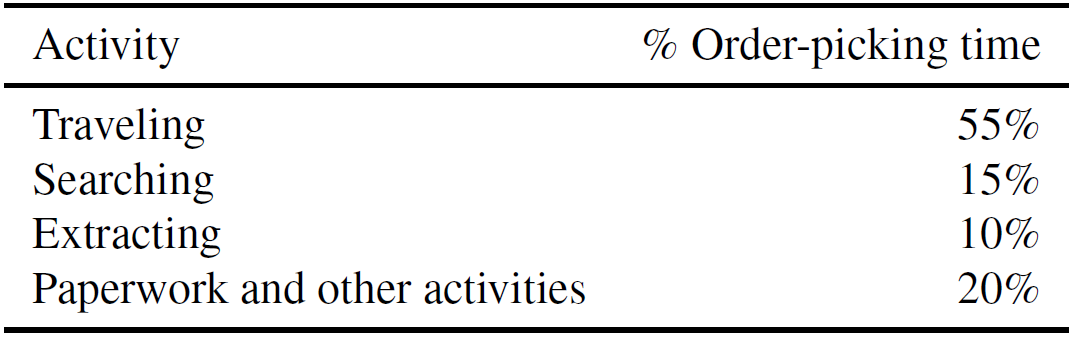}
\end{center}
\FloatBarrier

As traveling time has the greatest cost in order-picking, it is clear that majority of the design of the order-picking process is targeted in minimizing travel time in the warehouse. Any underperformance in the order-picking process may lead to unsatisfactory service for the customer and high operational costs for the warehouse itself, and ultimately the entire logistics network.

\section{Preliminaries}
In this section, we introduce some important assumptions and definitions with are key to defining our project objectives and formulating the optimization problems and heuristic algorithms. 

\subsection{The Warehouse}
We first state the following definitions about the warehouse and key terms which are widely used throughout the thesis:

\begin{definition}
The warehouse has a rectangular layout without unused space and only has parallel aisles. It contains a single depot used to take the order and to drop it off, and is also divided into blocks, which contains slots that store products, and are separated by cross-aisles. Cross-aisles do not contain any products but allow the picker to navigate in the warehouse. 
\end{definition}

\begin{remark}
Note that the warehouse must have at least one cross-aisle at the top and bottom, possibly containing more. 
\end{remark}

\begin{definition}
A subaisle is defined as a section of an aisle within a block. 
\end{definition}

\begin{figure}
\centering
\begin{minipage}{0.45\textwidth}
\centering
\includegraphics[width=0.9\textwidth]{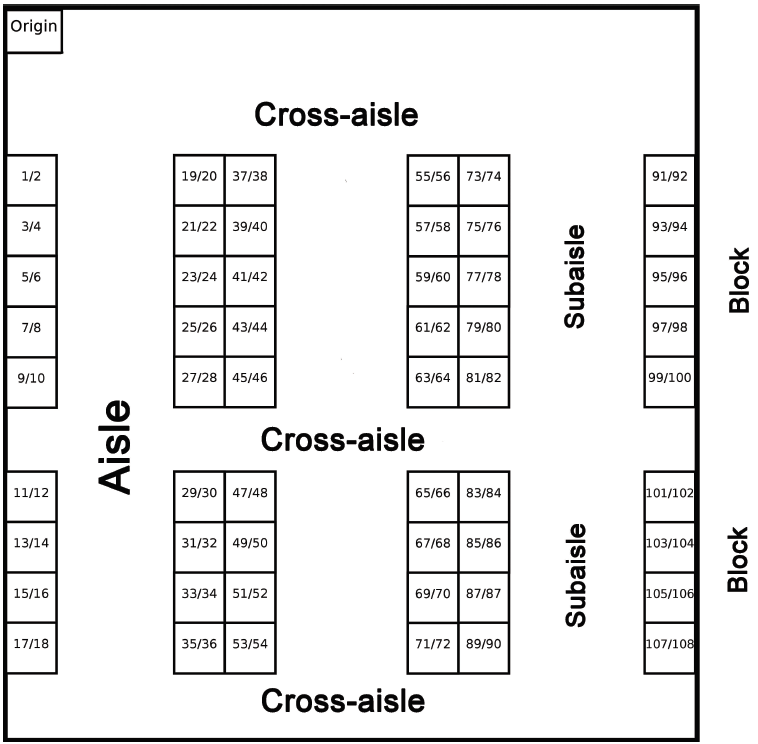}
\caption{Warehouse structure~\cite{valle2017}}
\label{warehouse_structure}
\end{minipage}\hfill
\begin{minipage}{0.45\textwidth}
\hspace*{-1cm} 
\includegraphics[width=1.2\textwidth]{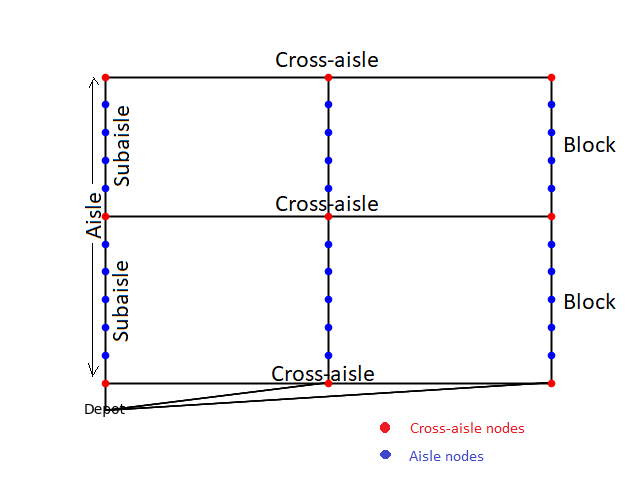}
\caption{Our representation of the warehouse as a graph}
\end{minipage}
\end{figure}

\begin{example}
A typical warehouse layout can be found in Figure \ref{warehouse_structure}, where there are 3 aisles, each aisle containing 2 subaisles.
\end{example}

\begin{definition}
An $\textbf{order picker}$ or $\textbf{picker}$ is a warehouse employee that is tasked with order picking.
\end{definition}

To collect products in the warehouse, the picker uses a picking device, which vary across warehouses, but it usually comes in the form of a cart/trolley, or a motorized vehicle with a certain storage capacity.

\FloatBarrier

Without loss of generality, we make the following assumptions about warehouse operations, which help simplify the explanation of the problems at hand:

\begin{assumption}
All aisles have equal lengths. All cross-aisles have equal lengths.
\end{assumption}

\begin{assumption}
Aisles contain slots on both sides, which can be stacked vertically in shelves and only one type of product is present in each slot.
\end{assumption}

\begin{assumption}
Pickers move along the center of the aisle.
\end{assumption}

\begin{assumption}
Products on both sides of the aisle are within reach of the picker.
\end{assumption}

In Figure \ref{full_warehouse_structure} below, slots with picks are marked in black. For illustration purposes, and also for the warehouse instances (based on real historical data) that we perform our experiments on, the depot is in front of the front aisle (also front cross-aisle of block 3) of the warehouse, aligned with the leftmost aisle of the warehouse. Note that in reality, the depot need not be at the position as shown in Figure \ref{full_warehouse_structure}, and can for e.g., lie anywhere along the front or rear aisle of the warehouse.
\begin{figure}[!htbp]
\centering
\includegraphics[scale=.65]{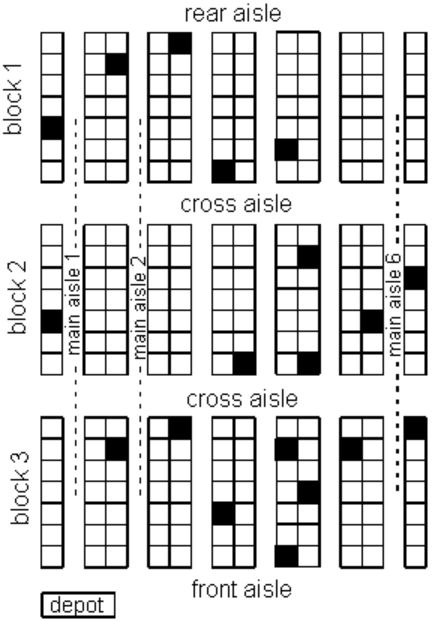}
\caption{Structure of a warehouse with pick locations~\cite{roodbergen2001}}
\label{full_warehouse_structure}
\end{figure}
\FloatBarrier

Throughout this thesis, we denote 

\begin{itemize}
\item $n_a = \text{number of aisles}$
\item $n_c = \text{number of }$ $\text{cross-aisles}$
\item $n_s = \text{number of shelves stacked vertically}$
\item $n_p = \text{number of distinct }$ $\text{products}$ 
\item $n_l$ the number of locations per aisle side (as a picker can pick from either side of the subsaisle)
\end{itemize}

Here, $n_l$ can be interpreted as the number of pallets in each shelf. As an example in Figure \ref{full_warehouse_structure}, we have $n_a = 6, n_c = 4, n_l = 21$. $n_s$ and $n_p$ is dependent on the order file data, which will be discussed in later chapters.

\subsection{The Problem Description}
Let $\mathcal{P}$ denote the set of products whose storage slots in the warehouse are known and $\mathcal{L}$ be the set of locations in the warehouse where a picker can pick up products. For example, these locations are the middle of an aisle containing products on both sides in different shelves, and from where the picker can reach these products. To be precise, each location $L \in \mathcal{L}$ contains a subset of products. 

\begin{definition}
An $\textbf{order}$ is given by a list of picks, i.e. a set of $|\mathcal{P}|$ products, indexed from $1$ to $m$ and described by their location in the warehouse. 
\end{definition}

Let $O$ denote the set of orders to be collected. Then each $o \in O$ contains a subset of products $P_o \subseteq \mathcal{P}$. Similarly, for each $o \in O$, the subset of locations $L_o \subseteq \mathcal{L}$ contains all products in $P_o$, and it is possible that multiple products from the same order are in a single location. We have $L(O) = \bigcup_{o\in O} L_o$ to represent the set of locations that contains all products that need to be picked from all orders. Also let $d_{lm} \geq 0$ be the distance between locations $l,m \in \mathcal{L}$ which are symmetric, such that $d_{lm} = d_{ml}$.

Let $\mathcal{T}$ denote the set of available pickers such that $T=|\mathcal{T}|$, $B$ be the number of baskets that a picker can carry and $b_o$ be the known number of baskets required to store all products in order $o\in O$. Here, we made the assumption that a basket will only consist of products from a single order, even if it is not filled to capacity. Lastly, let $s$ denote the depot (origin where pickers depart from and return to), and let $d_{sl}\geq 0$ be the distance between location $l$ and $s$.

The problem is stated as follows: given $|\mathcal{P}|$ products to pick in a rectangular warehouse, what is the minimum number of pickers required and the shortest tour (undertaken by each picker) which begins and ends at the depot to collect all these products?

The problem of routing each picker to get the shortest tour is a particular case of the Traveling Salesman Problem (TSP)~\cite{koster2007}, which is one of the most extensively studied problems in combinatorial optimization~\cite{scholz2016}, and it is probably the most notorious problem in Operations Research as it is very easy to explain, yet tempting to try and solve. The TSP is also $\mathcal{NP}-$hard~\cite{wolsey1998}. In the TSP, the salesman must visit each of the $n$ cities exactly once and then return to the origin. The objective is thus to find the order in which he should make his tour, so as to finish it as quickly as possible. In this thesis, the objective of the routing policy is to sequence the items on the list of picks, to ensure a good route through the warehouse.

Orders received by a warehouse can be rather large. In such cases, each order can be picked by a single picker. However, when orders are small, it is possible to reduce picker traveling times by grouping orders together subjected to a picker's carrying capacity, as lesser number of pickers are required, hence possibly resulting in an overall traveling time. Thus, we arrive at the following definition:

\begin{definition}
$\textbf{Order batching}$ is defined as the method of finding an optimal partitioning of a set of orders into a number of sub-sets, where each of these sub-sets can be retrieved by a single picker~\cite{koster2007}. Moreover, the problem is $\mathcal{NP}-$hard~\cite{gademann2005}.
\end{definition}

Let $\mathfrak{P}$ denote the set of all partitions of $O$, the set of orders. Then for $P\in \mathfrak{P}$, a partition $P$ of $O$ is feasible if the picker capacity is not violated, i.e. the total weight of all the orders with their items in each $p \in P$ does not exceed the picker's capacity $c$. Note that each $p$ is a set of orders, i.e. $p \subseteq O$. Let $d_{p}$ be the distance required to pick all items in $p$. Then we have $d_P = \sum_{p \in P} d_{p}$ to be the total distance covered across all $p \in P$. Let $w_{p}$ be the weight of each $p \in P$. The order batching problem is then to find a feasible partition with the minimum distance:

\begin{align}
\min\limits_{P \in \mathfrak{P}} \quad & d_P\\
\textrm{subject to} \quad & w_{p} \leq c, \quad \forall p\in P
\end{align}

\begin{example}
Suppose that the picker has a capacity of $8$ items and that we have 3 orders: $O = \{o_1, o_2, o_3 \}$, where the weights of the orders (in terms of number of items) are $w_1 = 4, w_2 = 6, w_3 = 4$ respectively. It is known that the distance required to pick all items in orders $o_1, o_2, o_3$ are $70, 92, 77$ respectively. Furthermore, the distance required to pick all items in the combined order $o_1 \cup o_3$ is $115$.  Thus we have $\mathfrak{P} = \{P_1 , P_2 \}$, where 
\begin{align*}
P_1 = \{ \{o_1 \} , \{o_2 \} , \{o_3 \} \} , P_2 = \{ \{ o_1 , o_3 \} , \{ o_2 \} \} 
\end{align*}
Since $d_{P_1} = 70+92+77 = 239$ and $d_{P_2} = 115 + 92 = 207$, we have the feasible partition with the minimum distance to be $ P_2 = \{ \{ o_1 , o_3 \} , \{ o_2 \} \} $.
\end{example}

\newpage

\subsection{Useful Resources}
Throughout this thesis, we illustrate some examples with images of a warehouse with predefined parameters. Those not cited are either generated with \textit{Interactive Warehouse}, publicly accessible at \url{http://www.roodbergen.com/warehouse/frames.htm}, or with \code{whopt.exe} software publicly available at \url{http://www.roodbergen.com/whopt/whopt.exe}. Routes from routing heuristics (to be introduced in later chapters) can also be drawn using these software.

\section{Summary of Contributions}
In this thesis, we demonstrate the impacts by various methods of solving with different batching and routing heuristics on the objective value and the computation time. In particular, we showed that in batching, having an optimal solver for routing of pickers does not always yield the lowest objective value. Methods of solving which utilize routing heuristics do result in a lower objective value as compared to the method where an optimal solver is used for picker routing. The second main contribution of this thesis is our recommendations made in answering the two main questions in our business problem, where we showed that the method of batching with routing heuristics and with routing of pickers with their batched orders solved optimally formed majority of the recommended methods of solving across all the warehouse instances. 

\end{doublespacing}

\chapter{Literature Review}

\begin{doublespacing}

\section{Heuristics}

\subsection{Background \& Applications}

In mathematical optimization, most of the practical problems that we wish to solve are $\mathcal{NP}$-hard. Heuristics or approximation algorithms are \textbf{techniques} designed to "solve" discrete optimization problems quickly. To be precise, they are optimization methods that attempt to make use of problem-specific information to obtain a high-quality solution for the problem and that there is no guarantee that they are able to find the optimal solution~\cite{rothlauf2011}. The main goal of heuristics is to obtain a "good" enough feasible solution quickly (e.g. an upper bound solution to an optimal solution of a minimization problem), which is usable in actual operations.

Heuristics are often crafted from information (i.e. problem-specific) about high-quality solutions and also takes into account a fixed set of rules. In most cases, it is noted that as the number of problems increases, the effectiveness of a particular method decreases. Thus, the effectiveness of the methods can be improved by narrowing down the problem and reducing the scope of its application. 

\subsection{Trade-offs}

Where there are benefits to utilizing heuristics, there are notable trade-offs when a heuristic is used instead of an optimal algorithm, and also between heuristics that are used for a problem instance. They are as follows:

\begin{itemize}
\item \textbf{Optimality:}
	\begin{itemize}
		\item Does the heuristic guarantee that the best solution will be found? 
		\item Must the best solution be found?
	\end{itemize}
\item \textbf{Completeness:}
	\begin{itemize}
		\item Is the heuristic able to find all the solutions (given that several exists)?
		\item Are all the solutions needed?
	\end{itemize}

\item \textbf{Accuracy and Precision:}
	\begin{itemize}
		\item Are the solutions obtained within a certain bound? For example, within a confidence interval.
	\end{itemize}
\item \textbf{Execution Time:} 
	\begin{itemize}
		\item Is this the heuristic with the quickest run-time?
	\end{itemize}
\end{itemize}
Given all these trade-offs, people implementing the heuristics will have to think hard about what is the end deliverable to for example, the business problem at hand. A consequence of these uncertainties is the narrowing down of the method of approach to a problem, and the experiments with several kinds of heuristics in an attempt to find the most suitable heuristic for each problem.

\section{The Picker Routing Problem}

\subsection{Exact Methods In Solving Picker Routing}
Various methods have been formulated and used to solve the TSP optimally. One well-known method is the Concorde~\cite{applegate2007} TSP solver, which is a program for solving the TSP. For the case of picker routing in warehouses, according to~\cite{valle2017}, the authors Ratliff and Rosenthal showed that the picker routing problem can be solved polynomially via a dynamic programming approach, for warehouses with a single block. Their algorithm was later generalized and the TSP in any series parallel graph was shown to be solvable in polynomial time. In particular, the TSP on sparse graphs was characterized as a Graphical TSP (GTSP). Unlike Ratliff \& Rosenthal, it is assumed that warehouses may contain multiple blocks. To the best of the authors' knowledge in~\cite{valle2017}, the GTSP in graphs made up of several connected series parallel subgraphs (e.g. warehouses with multiple blocks) has not been shown to be efficiently solvable. 

Let $d_{ij}$ be the distance between two points $i$ and $j$. Let also $N$ be the set of points (cities) and $S$ be a subset of cities that are to be traversed. Define
\[ x_{ij}=
	\begin{cases} 
      1 & \text{if salesman goes directly from town $i$ to town $j$,} \\
      0 & \text{otherwise.}
   \end{cases}
\]
The problem can then be formulated as the following, which will be passed to a solver (e.g. CPLEX) which will return optimal solutions:
\begin{align*}
\min \quad & \sum\limits_{i=1}^n \sum\limits_{j=1}^n d_{ij} x_{ij}\\
\textrm{subject to} \quad & \sum\limits_{j:j\neq i} x_{ij} = 1 \quad && \text{for $i=1,\dots,n$}\\
\quad & \sum\limits_{i:i\neq j} x_{ij} = 1 \quad && \text{for $j=1,\dots,n$}\\
\quad & \sum\limits_{i \in S} \sum\limits_{j \not\in S} x_{ij} \geq 1 \quad && \text{for $S \subset N, S \neq \emptyset$}\\
\quad & x_{ij} \in \{0,1\} \quad && \text{for $i=1,\dots,n$, $j=1,\dots,n$, $i \neq j$}
\end{align*}

\subsection{Heuristic Methods in Solving Picker Routing}
In actual practice, the problem of routing pickers in a warehouse to pick all items in the order allocated to them is mainly solved using heuristics. The main reasons~\cite{koster2007, gademann2005, roodbergen2001} why they are used instead of optimal algorithms are as follows: 

\begin{enumerate}
\item Not every warehouse is able to utilize an optimal algorithm. For example, a large warehouse instance with 8 aisles, 3 blocks and 33 possible pick locations per aisle is shown to not have any optimal solution within a time limit of 6 hours when trying to batch orders and route pickers optimally in ~\cite{valle2017}. Thus in reality, it may not be practical to use optimal algorithms due to possibly long computation times which may not even yield the required optimal solution within the day orders have been received by the warehouse. 
\item Optimal routes may seem illogical to pickers. As a result, they choose to deviate from the computed routes assigned to them. 
\item An optimal algorithm cannot take into account aisle congestion, whereas it is possible to avoid (or reduce) with routing heuristics. 
\end{enumerate}

One of the simplest and commonly used heuristics for picker routing is the S-shape heuristic. When routing pickers with this heuristic, the picker has to traverse aisles that contain at least one pick entirely, except possibly the last visited aisle. Aisles without picks are not entered at all. Once all aisles with picks have been visited, the picker returns to the depot. Another heuristic for picker routing is the Largest Gap heuristic, where a picker enters an aisle as far as the largest gap within an aisle and leaves each aisle from the same end. The gap is the separation between any two adjacent picks in the aisle, between the first pick and the front aisle, or between the last pick and the back aisle. If there is only one pick in the aisle, the picker picks and returns to the cross aisle it came from. 

The above heuristic methods were originally developed for single-block warehouses, but they can be used for multiple-block warehouses with certain modifications~\cite{koster2007}. Methods for routing pickers in multiple-block warehouses can be found in~\cite{roodbergen2001}, and will be explained in deeper detail in the next chapter of this thesis. 

\section{The Order Batching Problem}

\subsection{Exact Methods In Solving Order Batching}

The order batching problem with a general objective of minimizing the total travel time was shown to be $\mathcal{NP}-$hard by the authors in~\cite{gademann2005}. They employed a branch-and-price algorithm to solve instances of modest size (mostly warehouse instances with 10 \& 20 aisles and 20 to 30 pick locations per aisle side, and 15 to 32 number of orders) to optimality. In the case of larger instances, an iterated descent approximation algorithm was suggested. In~\cite{valle2017}, the authors formulate and solve the Joint Order Batching and Picker Routing Problem (JOBPRP), where the task is to find minimum-cost closed walks, where each picker visits all locations that allow the pickers to pick all products from their assigned orders. In their previous work, they formulated a directed model that involves exponentially many constraints to enforce connectivity requirements for closed walks. A branch-and-cut algorithm that relied on this non-compact model in their previous work was introduced They also examined the compact formulations in their previous work (which are based on network flows) using the CPLEX branch-and-bound solver. In~\cite{valle2017}, the authors focused on improving the non-compact formulation of the JOBPRP. They introduced several valid inequalities (cuts) based on a sparse graph representation of warehouses and showed that the introduction of the cuts greatly improved computational results. In particular, they show that when batching and routing problems are solved separately, optimal routing can be computed very quickly once all orders have been assigned to pickers. In this thesis, we used the integer linear programming (ILP) formulation for the JOBPRP introduced in~\cite{valle2017}, with constraints $(1)-(21)$.

\subsubsection{Notations}
In this thesis, we view the JOBPRP as a graph optimization problem like in~\cite{valle2017}. To define it, a directed and connected graph $D=(V,A)$ is introduced, where the set of vertices $V$ is given by the union of $s$, a set $V(O)$ containing a vertex for every $l\in L(O)$ and a set $V_A$ of artificial locations, which are located in corners between aisles and cross-aisles and do not contain products to be picked. Furthermore, we have the sets $V_o$ which contain a vertex for every $l\in L_o$, and $V(O)=\bigcup_{o\in O} V_o$. Hence, we have $V = \{s\} \cup V(O) \cup V_A$ and $|V| = 1 + |V(O)| + |V_A|$.

In the ILP formulation of the JOBPRP, vertices are allowed to be visited multiple times, but each arc cannot be traversed more than once. In particular, the formulation uses exponentially many constraints to enforce the connectivity of the closed walks~\cite{valle2017}. Let $z_{ot}$ indicate whether ($z_{ot} = 1$) or not ($z_{ot = 0}$) picker $t$ picks order $o\in O$, $x_{tij}$ to indicate whether ($x_{tij} = 1$) or not ($x_{tij} = 0$) arc $(i,j) \in A$ is traversed by trolley $t$, $\alpha_t$ to indicate whether ($\alpha_t = 1$) or not ($\alpha_t = 0$) picker $t$ picks at least one order and $y_{ti}$ to indicate whether ($y_{ti = 1}$) or not ($y_{ti} = 0$) vertex $i \in V \setminus \{s\}$ is visited by trolley $t$. Furthermore, we have $g_{ti} \in \mathbb{Z}_{+}$ to indicate the outdegree of vertex $i\in V$ in the closed walk for trolley $t$.

Let $\delta^{-} (W) = \{(i,j) \in A : i \not\in W, j \in W\}$ denote the set of inward-directed arcs, $\delta^{+} (W) = \{(i,j) \in A : i \in W, j \not\in W\}$ denote the set of outward-directed arcs, and $A(W) = \{(i,j) \in A : i \in W, j \in W\}$. Lastly, let $|\delta^{+} (i)|$ denote the maximum outdegree of $i\in V$. We are now ready to state the ILP formulation for the JOBPRP in the following section, where we used the first $21$ constraints for exact solving in our experiments.
\newpage

\subsubsection{The JOBPRP Formulation}

\begin{align}
&\min \quad 
\sum_{t\in\mathcal{T}} \sum_{(i,j)\in A} d_{ij} x_{tij} \label{objective}\\
\textrm{subject to}\quad
\sum_{o\in O} b_o z_{ot} &\leq B\alpha_t \quad && \forall t \in \mathcal{T} \label{2}\\
\sum_{t\in \mathcal{T}} z_{ot} &= 1 \quad && \forall o\in O \label{3}\\
\sum_{(i,j)\in \delta^+(i)} x_{tij} &\geq z_{ot} \quad && \forall o \in O, t\in \mathcal{T}, i:l(i)\in L_o\label{4}\\
\sum_{(i,j)\in \delta^+(i)} x_{tij} &= \sum_{(i,j)\in \delta^-(i)} x_{tji} \quad && \forall i\in V, t\in \mathcal{T}\label{5}\\
\sum_{(s,j)\in \delta^+(s)} x_{tsj} &= \sum_{(j,s)\in \delta^-(s)} x_{tjs} = \alpha_t \quad && \forall t\in \mathcal{T}\label{6}\\
x_{tij} &\leq \alpha_t \quad && \forall (i,j)\in A, t\in \mathcal{T}\label{7}\\
z_{ot} &\leq \alpha_t \quad && \forall o\in , t\in \mathcal{T}\label{8}\\
\sum_{o\in O} z_{ot} &\geq \alpha_t \quad && t\in \mathcal{T}\label{9}\\
\sum_{(i,j)\in \delta^+(i)} x_{tij} &= g_{ti} \quad && \forall i\in V, t\in \mathcal{T}\label{10}\\
y_{ti} &\geq x_{tij} \quad && \forall (i,j)\in A, t\in \mathcal{T}\label{11}\\
\sum_{j\in W} g_{tj} &\geq y_{ti} + \sum_{(j,k)\in A(W)} x_{tjk} \quad && \forall i \in W, W \subseteq V \setminus \{s\},|W|>1, t\in\mathcal{T}\label{12}\\
x_{tij} &\in \mathbb{B} \quad && \forall (i,j)\in A, t\in \mathcal{T}\label{13}\\
z_{ot} &\in \mathbb{B} \quad && \forall o\in O, t\in \mathcal{T}\label{14}\\
0 \leq \alpha_t &\leq 1 \quad && t\in\mathcal{T}\label{15}\\
y_{ti} &\leq \alpha_t \quad && \forall i\in V, t\in \mathcal{T}\label{16}
\end{align}

In the ILP formulation of the JOBPRP above, constraint $(2)$ ensures that the number of baskets (carted around by a picker on a trolley) does not exceed its given capacity and constraint $(3)$ ensures that each order is collected by exactly one picker. Constraint $(4)$ enforces the condition that if an order is assigned to a picker, then the vertex that stores a product of this order will be visited by the picker at least once. Constraint $(5)$ ensures that for every arc that leads to a vertex, there is one that departs from it. Constraint $(6)$ ensures that if a picker picks an order, it must depart from the origin $s$ (depot). Without significant loss of generality, it is assumed that the picker visits the source only once. Constraints $(7)$ and $(8)$ ensure that a picker visits an arc or picks an order only if it is used, while constraint $(9)$ ensures that if a picker is required, then at least one order is picked by the picker. Constraints $(10)$ and $(11)$ define the outdegree and $y_{ti}$ variables for each vertex $i$. According to~\cite{valle2017}, if the maximum outdegree of each vertex was not allowed to be greater than one but instead letting $g_{ti} \in \{0,1\}$, it would result in $y_{ti} = g_{ti}$, and constraint $(12)$ would change to the generalized subtour breaking constraint $\sum_{(j,k) \in A(W)} x_{tjk} \leq \sum_{j \in W\setminus \{i\}} y_{tj}$. In particular, constraint $(12)$ allows subtours found in closed walks as long as at least one vertex in the cycle has an outdegree of $2$. Constraints $(13)-(16)$ deal with the variables. The authors in~\cite{valle2017} noted that the $\alpha_t $ variables will be forced to binary even though constraint $(15)$ allows them to be fractional as this is due to the influence of other constraints (such as $(6)$ and $(13)$).

\subsubsection{Symmetry Breaking Constraints}
Symmetry breaking constraints were also introduced to break the symmetry in the space of feasible solutions. For example, Branch-and-bound algorithms based on symmetric formulations tend to perform poorly, as they enumerate search regions that lead to the same solution. Thus, the following constraint was added to the JOBPRP:
\begin{equation}
\sum\limits_{t=1}^o z_{ot} \geq 1, \quad o = 1,\dots ,T \label{17}
\end{equation}
(17) enforces that the first order goes to the first picker, the second order goes to either the first or second, and so forth. The following constraint was also added:
\begin{equation}
\alpha_t = 1, \quad t = 1,\dots , \ceil[\Big]{\sum\limits_{o \in O} b_o / B} \label{18}
\end{equation}
(18) ensures that the first minimum number of pickers are used.

Further symmetry breaking constraints were introduced as in order to break symmetry in directions adopted by each picker walk, then as distance is a symmetric function, we can enforce the condition that the arc out of $s$ for a picker is to the left (west) of the arc into $s$. The constraints which enforce this condition are:
\begin{equation}
\sum\limits_{k=a}^{W_A} x_{t,v(k,1),s} \geq \sum\limits_{k=a}^{W_A} x_{t,s,v(k,1)}, \quad \forall a \in \mathbb{A}\setminus \{1\}, t\in \mathcal{T} \label{19}
\end{equation}
\\
We also enforce the following constraints centered around the first cross-aisle vertex:
\begin{equation}
x_{t,v(a,1),S(a,1,1)} \geq x_{t,s,v(a,1)}, \quad \forall a \in \mathbb{A}, t\in \mathcal{T} \label{20}
\end{equation}
\begin{equation}
x_{t,S(a,1,1),v(a,1)} \geq x_{t,v(a,1),s}, \forall a \in \mathbb{A}, t\in \mathcal{T} \label{21}
\end{equation}

(20) ensures that ensures that the picker goes down the associated subaisle of the initial artificial vertex that he/she visits from the source, instead of going to another artificial vertex. (21) ensures that when the picker returns from an artificial vertex, he/she must have come from the associated subaisle(similar to Constraint 20).

The above formulation (JOBPRP) along with the additional constraints (17) - (21) were already implemented in MiniZinc by a previous intern (Joel) at Cosmiqo. Since the main scope of the project is to focus on heuristics, we will not be developing the JOBPRP any further due to time constraints and that the current formulation is sufficient to yield optimal solutions. 

\subsection{Heuristic Methods In Solving Order Batching}
Several heuristic methods to solve the order batching problem (as defined in the previous chapter) have been developed in recent decades. Notable methods include cluster analysis of orders to group them together~\cite{hwang2005}, seed-order selection rules~\cite{ho2008} and the widely-adopted Time Savings Heuristic~\cite{koster1999}. In the method of Cluster Analysis, similarity coefficients for all possible order pairs are computed and sorted, and order pairs are combined into a new order in order of decreasing similarity coefficients. For seed-order selection, to form an order batch, a seed order is first selected from the pool of orders using a seed-order selection rule. The selected seed order will be the first order added to the order batch, and updates will be made to the remaining capacity of the picker. Following which, another order selection rule is adopted to select another order from the pool of orders and add it to the batch of orders, while not exceeding the picker's capacity. This order selection process is repeated until the picker does not have any capacity for any more orders. In this thesis, we focus on the Time Savings Heuristic and implement it in our experiments, and we describe it as follows.

\subsubsection{Time Savings Heuristic}
The Time Savings Heuristic (TSH) \cite{koster1999} is used to compute batching of orders in the warehouse with a routing heuristic, that gives routing estimates used to compute the time (distance) savings. Let the time savings $s_{ij}$ be defined by
\begin{center}
	$s_{ij} = t_i + t_j - t_{ij}$
\end{center}
where $t_i, t_j$ are the order pick times for orders $i$ and $j$ respectively, and $t_{ij}$ is the order pick time of the order which consists of orders $i$ and $j$ combined. Both the S-shape and Largest gap routing algorithms are used.

\noindent The following algorithm is used to compute the time savings matrix and order batches:

\subsubsection{Basic variant, C\&W(i)~\cite{koster1999, clarke1964, elsayed1989}}
The algorithm consists of the following steps:
\begin{enumerate}
\item Calculate the savings $s_{ij}$ for all possible order pairs $i,j$.
\item Sort the savings in decreasing sequence.
\item Select the pair with the highest savings. If there is a tie, select a random pair.
\item Now, three cases can be distinguished:

	\begin{enumerate}
	\item Neither of the orders have been included in an existing route and the remaining capacity of the order picker is sufficient for both orders - include both orders in a new route.
	\item Exactly one order has been included in an existing route. If the other order fits in this route, add it to the route. If not, proceed with step 5.
	\item Both orders have been included in an existing route - go to step 5.
	\end{enumerate}

\item Select the next order combination from the list and repeat step 4 until all orders have been included in a route.
\end{enumerate}
If all order combinations have been selected, but not all orders have been included in a route: create a new route for every remaining order.

\end{doublespacing}

\chapter{Methods}

\begin{doublespacing}

\section{Our Approach to Solving the JOBPRP}

\subsection{Objectives}

The main objective of the project is to investigate the trade-offs when heuristics are used to obtain 'good enough' feasible solutions to the JOBPRP as compared to when the problem is solved to optimality with a solver. 

The methods employed (see later section) allow us to optimize the order picking process by batching several orders together, and by planning a good route (e.g. with heuristics) to minimize the distance required to pick up all the items in the orders. In particular, the proposed methods should scale up to typical warehouse sizes.

\subsection{Methods Employed In Experiments}

We adopted the ILP formulation of the JOBPRP with constraints $(1) - (21)$ from~\cite{valle2017}, along with Methods 1 to 3 in the paper in our experiments. 

Initially, we tried to solve the ILP formulation optimally. This method of using an ILP solver was shown to yield poor results (see later chapter on Results) in the form of sub-optimal solutions and extensive run-time of the solver used. As the ILP formulation is $\mathcal{NP}-$hard, we adopted batching and routing heuristics as an alternative to solving the ILP formulation optimally. For batching, it is computed via the Time Savings Heuristic (TSH) which also employs a routing heuristic to compute routing estimates, and for picker routing we employed the Nearest Neighbor, S-shape and Largest Gap heuristics. Thus, for each heuristic method of solving, there are two instances in which routing heuristics are used - in the TSH as well as routing if pickers after batching. In particular, we vary the routing heuristics used for each Method in both batching and routing of pickers after final assignment of orders.

To observe the benefits of replacing routing heuristics with optimal routing, we compare between Methods 1 to 3 in the subsections that follow, where we employ the Concorde~\cite{applegate2007} TSP solver for optimal routing.

\subsubsection{ILP Solver}
To obtain optimal solutions to the ILP formulation, we employed a branch-and-cut solver to solve the formulation written in an open-source constraint modeling language.

\subsubsection{Method 1}
In this method, we first compute the order batching via the time savings heuristic (TSH). Picker routing heuristics were used to obtain estimates of the partial route distances, for use in the TSH. Partial routes were computed by running each of the following heuristics: Nearest Neighbor, S-shape and Largest gap. To be precise, this method is used to obtain upper bound solutions to the JOBPRP, solely by the use of heuristics. 

\subsubsection{Method 2}
This method involves the use of a heuristic with optimal routing for the final assignment of orders. In other words, we use the routing estimates obtained during the routing algorithms in the TSH (to batch orders), but once all orders have been assigned (batched), we solve for each picker optimally to find its optimal route.

For our experiments, in the order batching stage of Method 2, we employ the TSH which uses the routing heuristics introduced in Method 1. Then once order batches are computed in Python 3, to solve for each picker's route optimally (with their batched order), we employ the Concorde TSP Solver. In particular, once orders have been batched to each picker, the JOBPRP is equivalent to the general TSP problem \cite{wolsey1998} as picker capacities have already been handled in the routing heuristics in the TSH. 

\subsubsection{Method 3}

Here, instead of using routing heuristics to compute the routing estimates in order batching and routing of pickers after final assignment of batched orders, we employ the Concorde TSP solver to compute optimal routes. 

\section{Routing Heuristics}

\subsection{Nearest Neighbor Heuristic}
The Nearest Neighbor (NN) algorithm was one of the first algorithms used to determine a solution to the traveling salesman problem. In it, the salesman starts at a random city and repeatedly visits the nearest city until all have been visited. It quickly yields a short tour, but usually not the optimal one.

The NN algorithm is easy to implement and executes quickly, but it can sometimes miss out on shorter routes which can be observed from a human perspective, due to its "greedy" nature. In the worst case, the algorithm results in a tour that is much longer than the optimal tour.
\vspace{1mm}\\
\begin{algorithm}[H]
\DontPrintSemicolon
\SetKwInOut{Input}{input}
\SetKwInOut{Output}{output}
\Input{A list of picks}
\Output{Sequence of nodes to be visited}
Set current position (start) to depot\;
\While{\text{there are picks not yet picked}}{
	Find the shortest path from the current position to the next closest unpicked item\;
	Traverse the shortest path to this pick and mark this pick as picked\;
	Set current position to the pick node\;
}
Return to depot\;
\caption{Nearest Neighbor}
\end{algorithm}

\subsection{S-shape Heuristic}
The S-shape heuristic~\cite{roodbergen2001} is a picker routing heuristic used to obtain partial route estimates for use in the TSH, and also to route all pickers individually once orders have been batched to each of them. The main idea of the S-shape heuristic is to skip all subaisles in which no picks are present, and any subaisle with at least one pick is traversed entirely. The following is an example of the S-shape routing\footnote{Retrieved from: http://www.roodbergen.com/warehouse/background.php}:

\begin{figure}[!htbp]
\centering
\includegraphics[scale=.78]{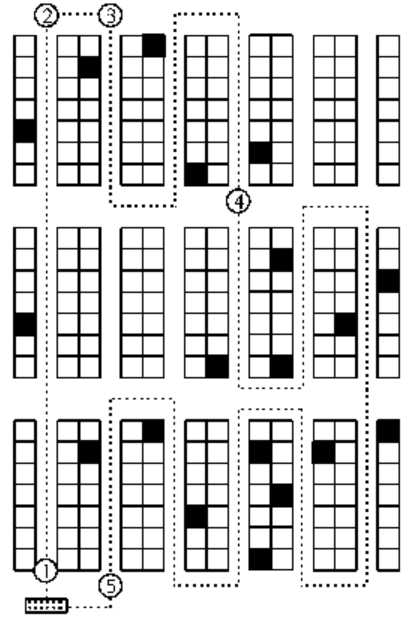}
\caption{A warehouse with order-picking performed by S-shape routing}
\label{s_shape_example}
\end{figure}
\FloatBarrier
The following is our implementation of the S-shape heuristic, built up upon the original implementation in~\cite{roodbergen2001}:
\vspace{1mm}\\
\begin{algorithm}[H]
\DontPrintSemicolon
\SetKwInOut{Input}{input}
\SetKwInOut{Output}{output}
\Input{A list of picks}
\Output{Sequence of nodes to be visited}
Retrieve all subaisles with at least 1 pick\;
Group all these subaisles (denote a subaisle as $sa$) with picks into blocks and sort the blocks in decreasing distance from the depot to get $B_s$\;
Determine the closest (left or rightmost) subaisle with pick in the furthest block from depot\;
Traverse the shortest path $P$ from the depot to the front of this subaisle and then to the furthest pick in the subaisle\;
Remove all picks from all blocks that lie on this shortest path to get resulting blocks $B_s^{(1)}$\;
Remove all subaisles in $B_s^{(1)}$ that no longer have any picks to get resulting blocks $B_s^{(2)}$\;
$node\_sequence \leftarrow P$\;
\caption{S-shape}
\end{algorithm}
\hspace{1mm}
\begin{algorithm}[H]
\DontPrintSemicolon
\setcounter{AlgoLine}{7}
\For{$b$ in $B_s^{(2)}$}{
Determine the closest $\mathit{sa}$ (left or rightmost) in the next block from the current position\;
    \If{closest $\mathit{sa}$ is rightmost $\mathit{sa}$ in next block}{
        Reverse the order of $b$\;
        }
    \# reverse the order of nodes in each odd-numbered $sa$ in each block\newline
    \For{$\mathit{sa}$ in $b$}{
        \uIf{$\mathit{sa}$ is even-numbered and not last subaisle}{
            $\mathit{node\_sequence} \leftarrow \mathit{node\_sequence} + \mathit{sa}$\;
        }
        \uElseIf{$\mathit{sa}$ is even-numbered and last subaisle}{
	     $\mathit{node\_sequence} \leftarrow \mathit{node\_sequence} + \textit{path to last pick in sa}$\;
	 }
	 \Else{
            $\mathit{node\_sequence} \leftarrow \mathit{node\_sequence} + \mathit{reverse}(\mathit{sa})$\;
        }
        
    }
}
Traverse across all nodes in $\mathit{node\_sequence}$ \;
Return to depot\;
\end{algorithm}
\vspace{1mm}

We first describe our implementation of the S-shape heuristic. The closest (left or rightmost) subaisle with pick(s) in the block furthest from the depot is first determined. The picker initially traverses the shortest path from the depot to the furthest pick in the furthest block (with at least one pick) from the depot. At this point, there are two possibilities: (a). There are no picks left in the current block; (b). There is at least one pick remaining in the current block.

If there are picks still remaining in the current block, the picker traverses the remainder of the subaisle entirely and arrives at a cross-aisle node at the back cross-aisle of the current block (for e.g. the picker is at $\mathbf{(2)}$ in Figure \ref{s_shape_example}). The picker then traverses the cross-aisle to the back of the next closest subaisle with at least one pick and traverses it entirely (e.g. subaisle with back cross-aisle node $\mathbf{(3)}$ in Figure \ref{s_shape_example}). This process is repeated until there are no more picks left in the current block. 

When there are no more picks in the current block, the picker traverses the shortest path from the last pick in the current block to the closest (left or rightmost) subaisle with pick(s) in the next block that is closer to the depot (e.g. subaisle with back cross-aisle node $\mathbf{(4)}$ in Figure \ref{s_shape_example}). The above process of S-routing is repeated until there are no more picks left to be picked in the warehouse. At this point, the picker traverses the shortest path back to the depot.

\subsection{Largest Gap Heuristic}
The Largest Gap heuristic is largely similar to the S-shape heuristic, with the exception of blocks being partitioned into front and back half-blocks by the largest gap between two adjacent picks in each subaisle. 

The following is an example of the Largest Gap Heuristic in action\footnote{Retrieved from: http://www.roodbergen.com/warehouse/background.php}:
\begin{figure}[!htbp]
\centering
\includegraphics[scale=.7]{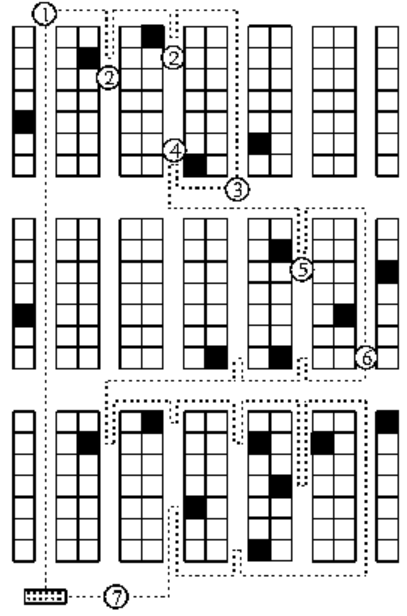}
\caption{A warehouse with order-picking performed by Largest Gap routing}
\label{largest_gap_example}
\end{figure}
\FloatBarrier

We define $d_1$ (resp. $d_2$) to be the total distance traversed by the picker when picking all picks from the front (resp. back) of the subaisle, and $d_3$ to be the maximum of all possible differences between the subaisle length and the gap between any two adjacent nodes in the subaisle. Let the minimum of the distances be denoted as $\min_d = \min \{d_1, d_2, d_3\}$. Thus, the process of partitioning each block by the largest gap is carried out by the following conditions for each subaisle:
\begin{enumerate}
\item If $\min_d = d_1$, add the subaisle to the list of front half-blocks.
\item If $\min_d = d_2$, add the subaisle to the list of back half-blocks.
\item Else partition the subaisle into front and back subaisles and add the half-subaisles into the respective front and back half-blocks.
\end{enumerate}

\noindent Our implementation of the Largest Gap heuristic introduced in~\cite{roodbergen2001} is as follows:
\vspace{3mm}\\
\begin{algorithm}[H]
\SetAlgoLined
\DontPrintSemicolon
\SetKwInOut{Input}{input}
\SetKwInOut{Output}{output}
\Input{A list of picks}
\Output{Sequence of nodes to be visited}
Retrieve all subaisles with at least 1 pick\;
Group all these subaisles (denote a subaisle as $sa$) with picks into blocks and sort the blocks in decreasing distance from the depot to get $B_s$\;
Determine the closest (left or rightmost) subaisle with pick in the furthest block from depot\;
Traverse the shortest path $P$ from the depot to the front of this subaisle and then to the furthest pick in the subaisle\;
Remove all picks from all blocks that lie on this shortest path to get resulting blocks $B_s^{(1)}$\;
Remove all subaisles in $B_s^{(1)}$ that no longer have any picks to get resulting blocks $B_s^{(2)}$\;
\# Partition each subaisle in $B_s^{(2)}$ into front and back subaisles, then combine all nonempty back (resp. front) sa together to get $B_s^{(3)}$, a list of half-blocks\newline
$B_s^{(3)} \leftarrow [\hspace{1mm}]$\;
\caption{Largest Gap}
\end{algorithm}
\hspace{1mm}
\begin{algorithm}[H]
\DontPrintSemicolon
\setcounter{AlgoLine}{7}
\For{$b$ in $B_s^{(2)}$}{
	$\mathit{front\_b} \leftarrow [\hspace{1mm}]$\;
	$\mathit{back\_b} \leftarrow [\hspace{1mm}]$\;
	\For{$\mathit{sa}$ in $b$}{
		$d_1 \leftarrow \textit{total distance when picking all picks from the front of } \mathit{sa}$\;
		$d_2 \leftarrow \textit{total distance when picking all picks from the back of } \mathit{sa}$\;
		\If{length$(sa)$ $> 1$}{
			Calculate gap between all adjacent node pairs in $\mathit{sa}$\;
			$d_3 \leftarrow \max (\textit{subaisle length} - \textit{gap between 2 adjacent nodes})$\;
		}
		$\min_d = \min \{d_1, d_2, d_3\}$\;
		\uIf{$\mathit{min_d} = d_1$}{
			$\mathit{front\_b} \leftarrow \mathit{front\_b} + \mathit{sa}$\;
		}
		\uElseIf{$\mathit{min_d} = d_2$}{
			$\mathit{back\_b} \leftarrow \mathit{back\_b} + \mathit{sa}$\;
		}
		\Else{
			$\mathit{front\_b} \leftarrow \mathit{front\_b} + \mathit{front\_sa}$\;
			$\mathit{back\_b} \leftarrow \mathit{back\_b} + \mathit{reverse} (\mathit{back\_sa})$\;
		}
	}
	$B_s^{(3)} \leftarrow B_s^{(3)} + [\textit{back\_b}, \textit{front\_b}]$\;
}
$\mathit{node\_sequence} \leftarrow P$\;
\end{algorithm}
\hspace{1mm}
\begin{algorithm}[H]
\DontPrintSemicolon
\setcounter{AlgoLine}{30}
\For{$b$ in $B_s^{(3)}$}{
	$\textit{current position} \leftarrow \textit{last node in node\_sequence}$\;
	Find the distances to the leftmost and rightmost $sa$ in $b$ from $\textit{current position}$\;
	\If{closest $sa$ is rightmost $sa$ in $b$}{
		$\mathit{reverse}(b)$\;
	}
	$\mathit{node\_sequence} \leftarrow \mathit{node\_sequence} + b$\;
}
Traverse all the nodes in $\mathit{node\_sequence}$\;
Return to depot\;
\end{algorithm}
\vspace{3mm}
We illustrate our implementation of the Largest Gap heuristic with the example from Figure \ref{largest_gap_example}. The picker traverses the shortest path from the depot to the furthest pick in the closest (left or rightmost) subaisle with pick(s) in the furthest block from the depot. With the remaining picks in the warehouse, partition each block into front and back half-blocks by the largest gap. At this point, we have a list of half blocks in order of blocks from the furthest to the closest to the depot. 

The picker now finds the next closest back subaisle (from the back half-block of the furthest block) and traverses the shortest path to the back cross-aisle node of this subaisle, picks all items from the back and then returns to the back cross-aisle node that he came from. This process is repeated until there are no more picks left in the back half-block (e.g. $\mathbf{(1) \rightarrow (2) \rightarrow (3)}$). Note that if initially there are no picks in the back half-block, the picker finds the closest (left or rightmost) front subaisle with pick(s) and traverses the shortest path to the front of this subaisle, picks all items from the front and returns back to the front cross-aisle node, repeating the process until there are no picks left to be picked in the front half-block. 

Once there are no picks left to be picked in the current block, the picker traverses the shortest path to the closest back subaisle with pick(s) in the next block. The process in the furthest block is repeated for this block and subsequent blocks until all picks in the warehouse have been picked. At which point, the picker traverses the shortest path back to the depot.

\section{Batching Heuristics}

\subsection{Time Savings Heuristic}

\begin{definition}
We define a router to be a function that takes in an order to be completed and computes a tour required to pick all items in the order. In particular, we define a savings (resp. batch) router as a function $\mathit{sr}(\mathit{order})$ (resp. $\mathit{br}(\mathit{order})$) that takes as input an order (a set of items), and outputs a tour that contains all the items in the order.
\end{definition}

\begin{remark}
Batches are a partition of orders.
\end{remark}

Let $s_{ij} = d_i +d_j - d_{ij}$, where $d_i = \mathit{distance}(\mathit{sr}(i)), d_j = \mathit{distance}(\mathit{sr}(j))$ are the distances required to pick order $i$ and $j$ respectively, and $d_{ij} = \mathit{distance}(\mathit{sr}(i \cup j))$ is the distance required to pick up the order where orders $i$ and $j$ are combined.

In our project, we parameterized the TSH with 2 parameters as input, a savings and batch router. In computing savings, a routing heuristic which is denoted as a savings router, is used to compute the routes (resp. distance traveled) for each picker which is used in the computation of savings. Once all orders have been batched, a batch router which is a routing heuristic is used to compute the route and distance for each batch. We implemented the TSH with the pseudocode:
\vspace{2mm}\\
\begin{algorithm}[H]
\SetAlgoLined
\DontPrintSemicolon
\SetKwInOut{Input}{input}
\SetKwInOut{Output}{output}
\Input{A list of orders, a savings \& batch router}
\Output{Sets of routes computed for each batched order}
$\mathit{savings} \leftarrow \{\hspace{1mm}\}$\;
\For{orders $i=1,...,n, j=1,...,n, i \neq j$}{
    \If{$\mathit{weight}(i) + \mathit{weight}(j) \leq \textit{picker capacity}$}{
        $s_{ij} \leftarrow d_i + d_j - d_{ij}$\;
        $\mathit{savings} \leftarrow \mathit{savings} \cup \{s_{ij}\}$\;
    }
}
Sort $\mathit{savings}$ in decreasing sequence\;
$\mathit{batches} \leftarrow \{\hspace{1mm}\}$\;
\caption{Time Savings}
\end{algorithm}
\hspace{1mm}
\begin{algorithm}[H]
\DontPrintSemicolon
\setcounter{AlgoLine}{9}
\For{$s_{ij}$ in savings}{
    \If{$i$ and $j$ $\notin$ batches}{
        $\mathit{batches} \leftarrow \mathit{batches} \cup \{i,j\}$\;
    }
    \If{$i$ $\in$ batches and $j$ $\notin$ batches and $\mathit{weight}(\mathit{batch_i}) + \mathit{weight}(j) \leq \textit{picker capacity}$}{
            $\mathit{batch_i} \leftarrow \mathit{batch_i} \cup \{j\}$\;
    }
    \If{$i$ $\notin$ batches and $j$ $\in$ batches and $\mathit{weight}(i) + \mathit{weight}(\mathit{batch_j}) \leq \textit{picker capacity}$}{
            $\mathit{batch_i} \leftarrow \mathit{batch_j} \cup \{i\}$\;
    }
}
$\mathit{routes} \leftarrow \{\hspace{1mm}\}$\;
$\mathit{distances} \leftarrow \{\hspace{1mm}\}$\;
\For{batch in batches}{
    $\textit{batch route} \leftarrow \mathit{br}(\mathit{batch})$\;
    $\mathit{routes} \leftarrow \mathit{routes} \cup \{\textit{batch route}\}$\;
    $\mathit{distances} \leftarrow \mathit{distances} \cup \{\mathit{distance}(\textit{batch route})\}$\;
}
Return $\mathit{routes}$, $\mathit{distances}$
\end{algorithm}
\vspace{2mm}
\begin{example}\label{tsh_example}
Suppose that we have seven orders, each consisting of five different items to be picked. In Figure \ref{savings_warehouse} below, the warehouse has seven aisles, with 15 item (resp. pick) locations per rack (resp. aisle). Here, the aisle length in 15m, with a crossover distance (to next adjacent aisle) of 4m, and the depot is located at the center of the first aisle. Assume for this example that the picker's capacity is eight items maximum, regardless of their actual weight. We define the weight of order $i$ as $w_i$, which corresponds to the number of items in that order. Suppose that there are seven orders, where $w_1 = 4, w_2 = 6, w_3 = 4, w_4 = 2, w_5 = 3, w_6 = 5, w_7 = 1$.  We now illustrate how TSH works based on the following warehouse instance which we generated\footnote{We used the Interactive Warehouse application created by Kees Jan Roodbergen: http://www.roodbergen.com/warehouse/frames.htm} using the above warehouse parameters:
\FloatBarrier
\begin{figure}[!htbp]
\centering
\includegraphics[width=0.52\linewidth]{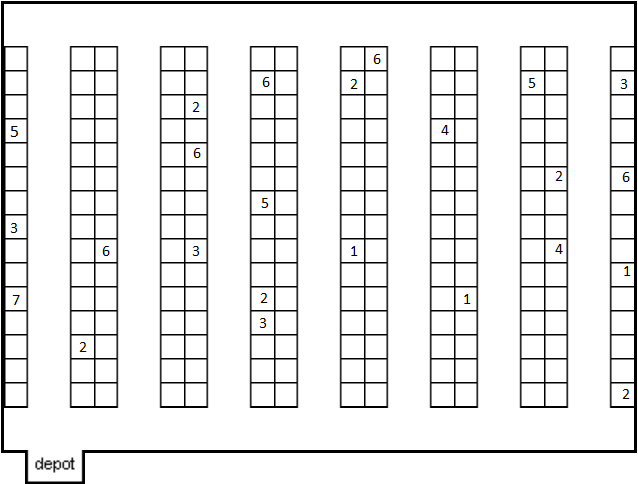}
\caption{An illustration of a warehouse with a block containing 7 aisles, with all items indicated with their order number.}
\label{savings_warehouse}
\end{figure}
\FloatBarrier

\begin{table}[!htbp]
\begin{center}
\begin{tabular}{c|ccccccc|c}
\hline
Order $i$ & 1 & 2 & 3 & 4 & 5 & 6 & 7 & Weight $w_i$ \\ \hline
1 & - &  &  &  &  &  &  & 4 \\
2 & X & - &  &  &  &  &  & 6 \\
3 & 59 & X & - &  &  &  &  & 4 \\
4 & 59 & 59 & 75 & - &  &  &  & 2 \\
5 & 67 & X & 94 & 54 & - &  &  & 3 \\
6 & X & X & X & 78 & 67 & - &  & 5 \\
7 & -10 & 9 & 9 & -4 & 9 & -10 & - & 1 \\ \hline
\end{tabular}
\end{center}
\caption{An example of a savings (in s) matrix}
\label{savings_example}
\end{table}
\FloatBarrier
Let the pick locations be as shown in Figure \ref{savings_warehouse}. Then we have the savings matrix as shown in Table \ref{savings_example}, where each entry in the matrix corresponds to the savings $s_{ij}$ for any order $i,j$, and with the respective order weights. Note that the savings matrix is symmetric, and an X indicates that the order combination is not possible due to the limitation of the picker's capacity. In this example, the savings router used is S-shape. The savings algorithm then proceeds in the following steps:

\begin{enumerate}
\item Select the order pair $\{3,5\}$, which has the largest savings. Since neither order is in an existing route and their total weight is $w_3 + w_5 = 7 < 8$, we cluster them together in a single route.
\item The order pair with the next largest savings is $\{4,6\}$ with a savings of $78$. Similarly, as neither order is in an existing route and their total weight is less than the picker's capacity, we cluster them in a single route.
\item The next largest savings is $75$, which corresponds to $\{3,4\}$. However, orders $3$ and $4$ are already in an existing route. 
\item The next largest savings is $67$, which corresponds to $\{1,5\}$ and $\{5,6\}$. We first select $\{1,5\}$, which has a total weight of $7$ and consists of order $5$, which is already in an existing route. As the total weight of the route containing orders $1,3,5$ is larger than the picker's capacity, it is not possible to add order $1$ to the existing route. Similarly, the order pair $\{5,6\}$ cannot be added to any existing route as both orders $5$ and $6$ are already in existing routes. 
\item The next possible pair is $\{2,7\}$. As neither of these orders are in an existing route, and that their total weight is $w_2 + w_7 = 7 < 8$, we cluster them into a new route. 
\item At this point, the order remaining is order $1$, which has not been added to any existing route. Since it cannot be added to any existing route due to pick capacity reasons, we create a new cluster for it. As a result, we have the resulting clustering of orders: $\{ \{3,5\}, \{4,6\}, \{2,7\}, \{1\} \}$.
\item Once clustering of orders have been computed, for each cluster $c_k$, we use the batch router to compute the route for each cluster by invoking: $\mathit{br}(c_k)$.
\end{enumerate}

\end{example}

\section{Summary of Methods}

In this thesis, we define the TSH to have two parameters: Savings Router \& Batch Router. Here, Savings Router denotes either a routing heuristic (Nearest Neighbor, S-shape, Largest Gap) or an Optimal router (Concorde TSP Solver).

In summary, we have the following table that shows the various combinations across Methods 1 to 3:

\begin{table}[!htbp]\centering\scriptsize
\begin{tabular}{|c|c|c|c|}
\hline
 & Method 1 & Method 2 & Method 3 \\ \hline
Savings Router & Nearest Neighbor, S-shape, Largest Gap & Nearest Neighbor, S-shape, Largest Gap & Optimal \\ \hline
Batch Router & Nearest Neighbor, S-shape, Largest Gap & Optimal & Optimal \\ \hline
\end{tabular}
\caption{Breakdown of combinations of routers used in each method}
\label{overview_of_methods}
\end{table}

\end{doublespacing}

\chapter{Experiments}

\begin{doublespacing}

\section{Objectives}
The main overall objective of the project is to investigate the trade-offs when heuristics are used to obtain 'good enough' feasible solutions to the JOBPRP as compared to when the problem is solved to optimality with a solver. The (heuristic) methods employed allow us to optimize the order picking process by batching several orders together, and by planning a good route to minimize the distance required to pick up all the items in the orders. 

In particular, we want to answer the following questions raised in the first chapter:

\begin{enumerate}
\item Across all order and warehouse instances, what is the quality of solution vs time trade-off?
\item If I have a warehouse with $X$ aisles and $Y$ cross-aisles and a number of orders to batch, what method should be used to solve the problem within the time limit?
\end{enumerate}

To answer the first question above, we show and discuss the results of median quality of solution and median time across all methods of solving for each warehouse instance, in the form of box plots for the quality of solution and tables for both median objective value and time elapsed. Furthermore, we only include the exact solving results for small warehouse instances as the large ones yielded very little results and the solutions obtained were all sub-optimal. 

As for the second question, it follows from the first in the form of recommendations and deeper analysis, where we also compare the spread of the data (quality of solution) on top of the medians. 

\section{Implementation}

\subsection{Routing Heuristics}
In this section of the thesis, we introduce our implementations of the routing heuristics in the following order: Nearest Neighbor, S-shape, Largest Gap, and briefly describe how we upgrade the simplified versions of the heuristic to that of the original in~\cite{roodbergen2001}. Once they were set up, we began working on the Optimal router for use in Methods 2 and 3.
\newpage

\subsubsection{Nearest Neighbor}
In this thesis, we employed the Nearest Neighbor, S-shape and Largest Gap heuristics for picker routing. Initially we began with the Nearest Neighbor heuristic as an initial naive implementation to observe how routing takes place in our experiments, and also as a soft start in configuring the Time Savings Heuristic (TSH), a batching heuristic which uses a routing heuristic to compute routing estimates used in the algorithm. Note that in practice, the Nearest Neighbor heuristic is not used for picker routing as the routes may turn out as illogical to pickers, and the heuristic may also result in 'bad routing' in some worst-case scenarios. For example, in a warehouse with 2 blocks with all but 1 pick in the block closest to the depot (i.e. 1 pick in furthest block), the picker may end up picking up all items in the block closest to the depot and then traverse the shortest path to the last pick in the furthest block. In this case, the picker may end up traversing along paths he/she has already traversed on to get to the last pick. This results in much additional distance traveled (resp. time). A snippet of the router class that employs the Nearest Neighbor algorithm is as follows:
\begin{python}
shortest_path = [start]
total_distance_traveled = 0
        
while unvisited:
    lengths, shortest_paths = 
                   shortest_path_to_all_nodes(shortest_path[-1], 
                                   self.shortest_paths, self.G)
            
    # get the shortest path to unvisited node with min distance
    closest_unvisited = min(unvisited, key=lengths.get)
    shortest_path_with_min_dist = 
                 shortest_paths[closest_unvisited]
    shortest_path += shortest_path_with_min_dist[1:]
    total_distance_traveled += lengths[closest_unvisited]
    unvisited.remove(closest_unvisited)
\end{python}
Here, \code{unvisited} is the list of picks in the order that are yet to be picked.

In our experiment setups, we used NetworkX\footnote{See documentation at: https://networkx.github.io/documentation/stable/}, which is a Python package for the creation, manipulation, and study of the structure, dynamics, and functions of complex networks. Initially for the Nearest Neighbor heuristic, in computing shortest paths from a pick node to another, we computed them repeatedly using \code{dijkstra\_path(G, source, target[, weight])}, which returns the shortest path from source to target in a weighted graph G as the warehouse can be treated as a weighted graph. This resulted in a long computation time for a single order, spanning as long as over 4000 seconds in method (i) for an order in a warehouse instance with 8 aisles, 2 blocks and 33 pick locations per aisle. We improved the computation time significantly to under 10 seconds for all orders of that warehouse instance by computing the shortest path to all nodes from each pick node and then storing these shortest paths in a dictionary. This was computed with \code{single\_source\_dijkstra(G, source[, target, ...])}, which compute shortest paths and lengths in a weighted graph G. We first import the NetworkX library as \code{nx}. The following is an example of our implementation of the function to compute shortest paths:

\begin{python}
def shortest_path_to_all_nodes(position, shortest_paths, G):
    if position not in shortest_paths:
        shortest_paths[position] = 
               nx.single_source_dijkstra(G, position)
    return shortest_paths[position]
\end{python}

In this function, \code{shortest\_paths} is the dictionary which stores the shortest paths computed from the nodes. As a result, shortest paths need not be repeatedly computed and can be retrieved from the dictionary instead. Thus we also implemented this method of computing shortest paths for other routing heuristics.

\subsubsection{S-shape}
Once the Nearest Neighbor heuristic was implemented, we began working on implementing the S-shape and Largest Gap heuristics proposed in ~\cite{roodbergen2001}. We started with a simplified version of the S-shape and Largest Gap heuristics. In these simplified heuristics, we route the picker in a standardized left-to-right fashion. In other words, the picker travels to the leftmost subaisle in the furthest block from the depot initially, then begins his/her routing from left to right. For example, in the S-shape case, after completing his/her routing in a block, the picker will travel to the leftmost subaisle with pick in the next block. Note that the picker will be at the rightmost subaisle with pick in the current block before traveling to the leftmost subaisle with pick in the next block, and that in the actual implementation of S-shape, the picker travels to the closest subaisle with pick in the next block upon clearing all picks in the current block he/she is in. It is clear that this implementation tends to result in several possible worst case scenarios in routing, for example starting from the leftmost subaisle with pick(s) in the next block may not be the closest subaisle with pick(s) from the current position (a cross-aisle node in the simplified heuristic implementation). We used the implementation of a simplified S-shape heuristic as a base which we built up on to upgrade it to a working implementation of the S-shape heuristic as introduced in~\cite{roodbergen2001}.

We have the following snippet of a key portion of our implementation of S-shape in Python 3 which performs the routing with the S-shape:
\begin{python}
node_sequence = [curr_pos]
for b in reduced_pick_blocks:
    last_pos = node_sequence[-1]
    closest_sa_and_path = 
               self.get_closest_sa_from_pick(b, last_pos)
    closest_sa = closest_sa_and_path.closest_sa
    updated_b = self.reverse_subaisles_order(b, closest_sa)
    last_sa_in_b = updated_b[-1]
    for sa in updated_b:
        if updated_b.index(sa) 
            node_sequence += list(reversed(sa.sa))
        elif updated_b.index(sa) 
            node_sequence += sa.sa
        else:
            last_pick = sa.picks[-1]
            node_sequence += sa.sa[:last_pick.index+1]
\end{python}

\subsubsection{Largest Gap}
Like with the S-shape, we also first implemented a simplified version of the Largest Gap heuristic. Each block in the warehouse is partitioned into front and back half-blocks, which each contain either front or back subaisles. Here, routing is standardized to be from right to left for the back half-block, and left to right for the front half-block. In order words, the picker will pick all picks in the back subaisles from left to right, and then pick all picks in the front subaisles from right to left. This is the case for every block in the warehouse. The following is a code snippet of the key portion of our implementation that sequences the half-blocks together:
\begin{python}
for b in partitioned_blocks:
    if b:
        curr_pos = node_sequence[-1]
        lengths, shortest_paths = 
           shortest_path_to_all_nodes(curr_pos, 
                           self.shortest_paths, self.G)
        dist_to_leftmost_node = lengths[b[0][0]]
        dist_to_rightmost_node = lengths[b[-1][0]]
    
        if dist_to_leftmost_node > dist_to_rightmost_node:
            reversed_b = list(reversed(b))
            node_sequence += [node for sa in reversed_b 
                                    for node in sa]
        else:
            node_sequence += [node for sa in b for node in sa]
\end{python}

\subsection{Optimal Routing}
In our experiments, we first ran tests for Method 1 followed by Methods 2 and 3. Methods 2 and 3 uses the PyConcorde for optimal routing (after batching for Method 2 and for both batching \& picker routing after final assignment). A code snippet of the optimal router which employs PyConcorde is as follows:
\begin{python}
# creates lower triangular distance matrix for all nodes in H
distances = 
        self.make_lower_diag_string(nodes_in_H, distance_matrix)
.
.
with os.fdopen(fd, 'w') as output_file:
    # The following is the formatting of the .tsp file
    output_file.write("NAME : Concorde TSP Solver\n")
    output_file.write("COMMENT : Solver for Method (ii)\n")
    output_file.write("TYPE : TSP\n")
    output_file.write("DIMENSION : 
    output_file.write("EDGE_WEIGHT_TYPE : EXPLICIT\n")
    output_file.write("EDGE_WEIGHT_FORMAT : LOWER_DIAG_ROW\n")
    output_file.write("EDGE_WEIGHT_SECTION\n")
    output_file.write("
    output_file.write("EOF\n")
    output_file.write(distances)
.
.
# map the vertex numbers back to the ones in G
node_sequence = [nodes_in_H[i] for i in solution.tour]
objective_value = solution.optimal_value
\end{python}

Another important function in each router class is the \code{route} function, which executes the routing to pick all items in the order with a routing heuristic as specified by the router class. A snippet of the function that applies to all four routers is as follows:

\begin{python}
def route(self, order):
        start = 1 #depot
        end = 1 #depot
        .
        .
        .
        total_distance_traveled = 0
        for node in node_sequence:
            lengths, shortest_paths = 
               shortest_path_to_all_nodes(shortest_path[-1], 
                                  self.shortest_paths, self.G)
            shortest_path += shortest_paths[node][1:]
            total_distance_traveled += lengths[node]
        
        route_with_distance = 
              namedtuple('RouteAndDistance',
                         ['shortest_path', 'distance'])
        route_tuple = 
              route_with_distance(shortest_path,
                                  total_distance_traveled)
        
        return route_tuple
\end{python}
In the \code{route} function, the \code{node\_sequence} is the sequence of nodes to be traveled by the picker, which begins and ends at the depot. The output of the function is a namedtuple that stores the shortest path to pick all items in the order and the total distance traveled by the picker.

\subsection{Batching Heuristics}

\subsubsection{Time Savings}
As we began our experiments with Method 1 which uses the TSH for batching of orders, once we had a working version of the Nearest Neighbor router, we began working on the TSH. Recall that the TSH takes in a savings router and a batch router as input. The TSH will first compute the savings by employing the savings router in computing the routing estimates for each picker and their order. Once all orders have been batched, the TSH will employ a batch router (which may be different from the savings router depending on the method of solving) to compute the route for each batch. A snippet of the TSH implementation in Python 3 with key portions of the code highlighted is as follows:

\begin{python}
def time_savings_heuristic(orders,savings_router,batch_router):
    savings = calculate_savings(savings_router, orders)
    sorted_savings = sort_savings(savings)
    batches = 
       calculate_batches(savings_router, sorted_savings, orders)
    .
    .
    for i in range(len(order_batches)):
        order_batch_route = batch_router.route(order_batches[i])
        route_traversed = order_batch_route.shortest_path
        all_routes_traversed += [route_traversed]
        distance_traveled = order_batch_route.dist
    .
    .
    return tsh_results
\end{python}

\subsection{Exact Solving}
For solving the ILP formulation of the JOBPRP, we employed the Coin-or Branch and Cut\footnote{See CBC documentation at: https://projects.coin-or.org/Cbc} (CBC) solver (which is an open-source mixed integer programming solver written in C++), and ran it with the MiniZinc\footnote{MiniZinc is a free and open-source constraint modeling language. See https://www.minizinc.org/.} formulation of the JOBPRP.

\subsection{Input Preprocessing}
Before the methods of solving with heuristics can be performed, much preprocessing on the input data has to be done. This involves reading the raw order file data, generating the warehouse parameters in Python 3, and also formatting the warehouse parameters in a way such that they can be used as input for the algorithms. One key stage is that of reading the warehouse parameters as input into a function, which then creates subaisles that can be used efficiently by the various heuristics.

When the warehouse has more than one block, partitioning the aisles into subaisles becomes non-trivial as aisles may contain an odd number of pick locations, and thus not every subaisle has the same number of pick locations. If the number of pick locations per aisle is even, it follows that the number of pick locations per subaisle is just the number of pick locations per aisle divided by the number of blocks in the warehouse. 

We implemented an algorithm that constructs the subaisles in each warehouse instance in our experiments. Here we introduce new variables: aisles - a list of aisles where each aisle itself is a list of nodes belonging to that aisle in the warehouse, ca\_nodes - the list of cross-aisle nodes of the warehouse and num\_block - the number of blocks in the warehouse. With the notations of the warehouse parameters defined previously, we have the following pseudocode:
\vspace{2mm}\\
\begin{algorithm}[H]
\SetKwInOut{Input}{input}
\SetKwInOut{Output}{output}
\Input{$n_a$, aisles, ca\_nodes, $n_l$, num\_block}
\Output{A list of subaisles}
\DontPrintSemicolon
$min\_size\_of\_sa  \leftarrow \lfloor(n_l / \text{num\_block})\rfloor$\;
\# initialize array containing initial sizes of subaisles\\
$sa\_sizes \leftarrow [min\_size\_of\_sa] \times num\_block$\;
$remaining\_nodes \leftarrow n_l - min\_size\_of\_sa \times num\_block$\;
\# add remaining nodes from left to right of sa\_sizes, i.e. populate from sa closest to depot\\
\For{$i$ in range(remaining\_nodes)}{
        $sa\_sizes[i] \leftarrow sa\_sizes[i] + 1$\;
}
$ca\_nodes\_by\_aisle \leftarrow [\hspace{1mm}]$\;
\For{$i$ in range($n_a$)}{
    $ca\_nodes\_by\_aisle \leftarrow ca\_nodes\_by\_aisle [ca\_nodes[j] $ $\textit{ for j in range}(i, length(ca\_nodes), n_a)]$\;
}
$subaisles \leftarrow [\hspace{1mm}]$\;
\caption{Input Preprocessing For Generating Subaisles}
\end{algorithm}
\hspace{1mm}
\newpage
\begin{algorithm}[H]
\setcounter{AlgoLine}{13}
\For{$i$, aisle in enumerate(aisles)}{
        $ca\_pos \leftarrow 0$\;
        $step = i * (num\_block + 1)$\;
        \For{$j$, size in enumerate(\text{sa\_sizes})}{
            $subaisle = [ ca\_nodes\_by\_aisle[step+j] ] + aisle[ca\_pos : ca\_pos + size] + [ ca\_nodes\_by\_aisle[step+j+1] ]$\;
            $subaisles \leftarrow subaisles + [subaisle]$\;
            $ca\_pos \leftarrow ca\_pos + size$\;
}
}
Return $subaisles$\;
\end{algorithm}
A snippet of our implementation of the above algorithm in Python 3 is as follows:

\begin{python}
def generate_subaisles(num_pick_loc_per_aisle, num_block, 
                                 aisles, cross_aisle_nodes):
    min_size_of_sa = num_pick_loc_per_aisle // num_block
    
    # initialize array containing initial sizes of subaisles
    sa_sizes = [min_size_of_sa] * num_block
    remaining_nodes = num_pick_loc_per_aisle 
                    - min_size_of_sa * num_block

    # add remaining nodes from left to right of sa_sizes, 
    #i.e. populate from sa closest to depot
    for i in range(remaining_nodes):
        sa_sizes[i] += 1

    ca_nodes_by_aisle = []
    num_aisle = len(aisles)
    for i in range(num_aisle):
        ca_nodes_by_aisle += [cross_aisle_nodes[j] for j in 
                              range(i, len(cross_aisle_nodes), 
                              num_aisle)]

    subaisles = []
    for i, aisle in enumerate(aisles):
        ca_pos = 0
        step = i * (num_block + 1)
        for j, size in enumerate(sa_sizes):
            subaisle = [ ca_nodes_by_aisle[step+j] ] 
                       + aisle[ca_pos : ca_pos + size] 
                       + [ ca_nodes_by_aisle[step+j+1] ]
            subaisles += [subaisle]
            ca_pos += size
    return subaisles
\end{python}

\section{Warehouse \& Order Instances}

\subsection{Warehouse Representation}
According to~\cite{valle2017}, there is no information about warehouse layouts and product placement in Foodmart, so the authors constructed a warehouse layout generator in the Perl programming language to simulate both. The generator creates warehouses which must be able to hold a minimum predetermined number of distinct products ($n_p$ ) given a (fixed) number of aisles, cross-aisles and shelves. Arbitrary lengths in meters for the widths of aisles, cross-aisles, rack depth and slots are given. The distance from the depot $s$ (origin) to its closest artificial vertex (which lies in the cross-aisle closest to the depot) is also given. 

The generator computes the number of slots a shelf must have in order to hold at least the required number of products $n_p$, while minimizing the number of empty slots. It also computes the position of cross-aisles such that aisles are divided in subaisles as equally (in terms of number of slots) as possible. The placement of products in slots is performed by sorting all products from the highest category level to the lowest and placing them in consecutive slots, such that similar products are close to each other. 

In our experiments, we experimented on small and large instances of the warehouse. For small warehouse instances, they consist of $2$ to $8$ aisles, $1$ block and $4$ possible pick locations per aisle. For large warehouse instances, they consist of $8$ to $30$ aisles, $1$ to $4$ blocks and $33$ possible pick locations per aisle. We standardized each shelf to hold $33$ slots, so that each warehouse can store all distinct products, enough for all $1560$ products in the Foodmart database. This condition is enforced by setting the number of shelves stacked vertically to $3$. The distance from the origin to the closest artificial vertex (directly in-front of depot) is $4$m, the aisle and cross-aisle widths are $3$m, and both the slot width and rack depth are $1$m. 
\newpage

\subsection{Test Instances}

\subsubsection{Orders}
The authors in~\cite{valle2017} noticed that orders are generally very small, and decided to combine different Foodmart orders into a single one to produce larger orders. For every customer, all of their purchases made in the first $\Delta$ days are combined into a single order. It is noted that the combined order may contain not only more distinct products, but also a higher quantity of items of a single product. 

A test instance is thus taken as the $O$ orders with the highest number of distinct products. If $O = 8$, the $8$ largest combined orders make up the test set, and if $O = 9$, we take the same orders as in $O = 8$ plus the ninth largest combined order. In our experiments, each order file (test instance) was created from $\Delta = \{5,10,20\}$ and $O = \{5,\dots , 50\}$. All test instances are publicly available as mentioned in a previous section.

The order instances data are publicly available from the MySQL Foodmart Database: \url{http://pentaho.dlpage.phi-integration.com/mondrian/mysql-foodmart-database} and the warehouse generator (together with some test instances as described in \cite{valle2017}) can be found at: \url{https://homepages.dcc.ufmg.br/~arbex/orderpicking.html}. The database consists of anonymised customer purchases over two years, across a chain of supermarkets. In total, there are $1560$ unique products classified into $4$ categories. In particular, it contains $\approx 270000$ orders for the period $1997 - 1998$, where each order has a customer ID, a list of distinct products purchased, and the number of items for each distinct product and their purchase dates. 

\subsubsection{Capacity of Baskets \& Number of Pickers}
The authors in~\cite{valle2017} defined each basket to hold a maximum number of $40$ items, irrespective of their sizes and weights. For every test instance, they also defined the number of available pickers to be $T = \Big{\lceil} \frac{\sum_{o\in O} b_o }{B} + 0.2 \Big{\rceil}$. They did not tackle the problem of finding the exact minimum $T$ required to service all orders as it is an optimization problem on its own. Note that not all pickers need to be employed to be used as the solution may not utilize every picker. In our experiments, we experimented with various picker capacities $c \in \{80, 160, 240, 320\}$. For example, a picker capacity of $160$ corresponds to $4$ baskets. We noted that varying picker capacities yielded results which show a general trend (in increasing quality of solution) which can be easily explained by the fact that larger capacities imply a greater ability for orders to be combined and batched together and thus an overall lower objective value (which leads to higher quality of solution). As such, we decided to stick to a single picker capacity of $320$, equivalently in~\cite{valle2017} where the authors set the number of baskets to $8$. 

\subsubsection{Summary of Warehouse Parameters \& Picker Capacities}
The following table is a summary of all our input parameters for the warehouse and picker capacity, and also that of the methods of solving employed:

\begin{table}[!htbp]
\centering
\resizebox{\textwidth}{!}{%
\begin{tabular}{|c|c|c|c|c|c|c|c|c|}
\hline
\textbf{Warehouse} & \textbf{Capacity} & \textbf{Methods} & \textbf{Aisles} & \textbf{Blocks} & \textbf{Shelves} & \textbf{Products} & $\mathbf{\Delta}$ & $\mathbf{O}$ \\ \hline
Small & 80, 160, 240, 320 & 1, 2, 3, Exact & 2, 3, 4, 5, 6, 7, 8 & 1 & 3 & 1560 & 5, 10, 20 & 5, 6, \dots, 50 \\ \hline
Large & 320 & 1, 2, 3 & 2, 3, 4, 5, 6, 7, 8, 10, 20, 30 & 1, 2, 3, 4 & 3 & 1560 & 5, 10, 20 & 5, 6, \dots, 50 \\ \hline
\end{tabular}%
}
\caption{Summary of warehouse parameters used for test cases in our experiments}
\label{warehouse_parameters_summary}
\end{table}
\newpage

\section{Experimental Setup}

\subsection{Setting Up A Workflow}
Snakemake\footnote{See https://snakemake.readthedocs.io/en/stable/ for documentation.}, a workflow management system described in Python based language, is used to run experiments locally on our PCs, and to submit jobs to the cluster in NUS HPC. To be precise, all our experiments were ran by invoking Snakemake rules in Bash. An example of a Snakemake rule from our experiments is as follows, which runs a test for Method 1, with Nearest Neighbor as the router, an order file with $\Delta = 10, O = 10$, a warehouse with $8$ aisles, $2$ blocks, $3$ shelves, $1560$ minimum number of products and a picker capacity of 320:

\begin{python}
rule method1_nn_nn_d10ord10:
    output: 
        "method1_d10ord10_nn_nn_8_1_3_1560.csv"
    shell:
        "python3 time_savings_heuristic.py "
        "-g ../data/foodmart/warehouseGenerator.pl "
        "-l ../data/foodmart/productsDB_1560_locations.txt "
        "-o ../data/foodmart/order/instances_d10_ord10.txt "
        "-csv {output} -c 320 -na 8 -nc 1 -ns 3 -np 1560 "
        "-sr nn -br nn"
\end{python}

\subsection{Running The Experiments}
The main required libraries and packages in Python 3 are $\textit{networkx}$, $\textit{pandas}$, $\textit{matplotlib}$, $\textit{scikit-learn}$, and $\textit{seaborn}$. Using Python 3, we generate the warehouse instances (warehouse\_input\_reader.py), employ the savings \& routing heuristics (time\_savings\_heuristic.py, NearestNeighborRouter.py, etc), and also employ the PyConcorde solver (OptimalRouter.py) used in methods 2 \& 3. 

In the initial stages of the project, we performed experiments on the smallest possible instance of the warehouse with 2 aisles, 1 block and 2 pick locations per aisle in Python 3. These were performed with the TSH and Nearest Neighbor heuristic, and we invoked the Snakemake rule that runs the shell command to run the experiment. We also used Git\footnote{For more details about Git, see https://gitforwindows.org/} for Windows to store results in Cosmiqo's GitLab account's repository.

Before we began working on Methods 2 and 3, we installed a working version of PyConcorde (a Python wrap-around the Concorde TSP solver) from \url{https://github.com/jvkersch/pyconcorde} and imported it into our implementation of the optimal router. It was also installed in the HPC terminal so that we are able to run experiments for Method 3, which can take very long to solve for large warehouse instances.

At the same time, we also worked on invoking MiniZinc with Snakemake to run experiments for exact solving. The JOBPRP with constraints $(1)-(21)$ were set up in MiniZinc by Joel - a previous student working on the project, and solved to optimality with the CBC solver. With the MiniZinc driver on the environment PATH variable in the system, we were able to perform exact solving on the ILP formulation of the JOBPRP with CBC using the following Snakemake rule:

\begin{python}
rule instances_d5_ord5_exact:
   input: 
        "instances_d5_ord5.dzn"
   output: 
        "instances_d5_ord5_exact.exact"
   shell: 
        "minizinc --solver osicbc ../minizinc/valle2017.mzn "
        "{input} --output-time > {output}"
\end{python}
In this Snakemake rule, the input file is the \code{.dzn} file that contains the picker capacity, warehouse parameters and edges (with their weights). The file used in this example corresponds to the order file with $\Delta = O = 5$. In the shell, the CBC solver is invoked with the \code{--solver osicbc} flag, and the time elapsed for solving is printed with the \code{--output-time} flag. The line \code{> {output}} results in solver outputs being saved to the \code{instances\_d5\_ord5\_exact.exact} file.

Experiments were initially conducted locally on our PCs via Ubuntu\footnote{Ubuntu is a free and open-source Linux distribution based on Debian. For more details, see https://www.ubuntu.com/} by invoking Make and Snakemake rule commands, which allows us to run batches of jobs in one go. Throughout the project, we implemented other routing heuristics in Python 3 and increased the size of the warehouse instances by increasing the number of aisles and the number of pick locations per aisle, and also increased the size of the orders. We noticed that the computation times grew as the size of the warehouse and order instances increases. It eventually reached a point where it took more than a day to run all experiments for all 138 order files for a large warehouse instance. In particular, the solving times for each order and warehouse instance for large warehouse instances were so large and memory consuming that it became no longer feasible to run on our personal computers.

As the processing power of our computers are not sufficient in conducting the tests (especially for solving large instances to optimality and for Method 3), we eventually reached out to the university's Information Technology's High-Performance Computing\footnote{See https://nusit.nus.edu.sg/services/hpc/about-hpc/} (HPC) center, where it became possible to submit jobs to the cluster (using Snakemake) for running of these experiments. This was done by submitting a job script via the Snakemake rule:

\begin{python}
rule submit_job:
    shell: 
        "python3 -m snakemake --cluster qsub -j 30 "
        "--jobscript js.txt all_exact"
\end{python}
where the \code{-j 30} option limits the number of concurrently submitted jobs to the cluster at $30$, the job script submitted is \code{js.txt} and the Snakemake rule to run which contains the experiment(s) is \code{all\_exact}. An example of the job script we used for submission is as follows:

{\scriptsize
\begin{Verbatim}[frame=single]
#!/bin/bash

## -P Exact_HPCTMP: Job project name
#PBS -P Exact_HPCTMP

## -q Queue_Name: which queue to sbmit the job to in HPC
## Note: parallel12 has wall time of 720 hours. 
#PBS -q parallel12

## -l reserves 1 units of 1 cpus, 5GB memory for this job
#PBS -l select=1:ncpus=12:mem=5GB

## -j oe states to join output and error files together
#PBS -j oe

## -N Exact_Solving_Job_Output: set filename for standard output/error message.
#PBS -N Exact_HPCTMP_Job_Output

## Change to the working dir in the exec host
cd $PBS_O_WORKDIR;

##--- Put your exec/application commands below ---
## source /etc/profile.d/rec_modules.sh gets path where modules are installed
source /etc/profile.d/rec_modules.sh
module load python3.6.4
## permanently have MiniZinc driver on environment PATH in HPC
export PATH="$PATH:/home/svu/e0004335/minizinc/bin"
## exec_job is where snakemake inserts the command
{exec_job}
\end{Verbatim}
}

\subsection{Data Analysis}
Plotting of picker paths was done in Python 3, with the aid of the $\textit{matplotlib}$\footnote{See documentation at: https://matplotlib.org/} library alongside the $\textit{networkx}$ library. Statistical visualizations such as box plots were created with the import of the $\textit{seaborn}$\footnote{See documentation at: https://seaborn.pydata.org/} library. We created box plots to compare the scalability of the warehouse, the impact of partitioning the warehouse into multiple blocks on the objective value, and also to gather deeper and more direct insights on the difference in methods for recommendation purposes.

\subsection{Problems Encountered \& Improvements Made}
Most of the problems we encountered throughout the duration of the project is during the configuration of the HPC software and terminal, in order for our experiments to run correctly in the HPC system. During our first few weeks of configuring the terminal, we encountered various problems such as an outdated kernel in HPC when running exact solving with CBC in MiniZinc. As a result, the MiniZinc bundle had to be rebuilt on an older kernel for it to run in HPC. Other (minor) problems encountered were missing packages in the HPC system for Python 3, and Git Large File Storage (LFS) was also missing. This resulted in file transfers being done manually over a secure SSH client such as Filezilla. Another somewhat significant problem faced was the lack of memory for exact solving with CBC in HPC. Initially, the number of pickers available was preset to 40, which was well-beyond the minimum number of pickers required to pick all items in each order, across all order and warehouse instances. A large number of available pickers does increase the solver search time significantly as it directly affects several constraints in the ILP formulation of the JOBPRP, by increasing the search space of the solving algorithm. We eventually reached a solution to this by first computing the minimum number of pickers required from the results of experiments for Methods 1 to 3, and then exported the number of pickers to the \code{.dzn} file which contains the required variables and parameters used in exact solving. As a result, solving times and memory usage in the terminal were greatly reduced. 

\end{doublespacing}

\chapter{Results}

\begin{doublespacing}

\section{Overview \& Preliminaries}
In our experiments, we employ $2$ methods of solving - exact and heuristic solving. For exact solving, we employ the ILP formulation of the JOBPRP with constraints $(1) - (21)$ in MiniZinc, and utilize the CBC solver to solve to optimality. For heuristic solving, we employ Methods 1 - 3 as described in Chapter $3$. We also computed the results for when no batching heuristic was used, which we denote as trivial batching where each order is assigned to a picker and routing is solved with PyConcorde for each picker. To compare the result of using a batching heuristic and without (trivial batching), we compute the quality of solution when a batching heuristic is used with the following formula:

\begin{definition}
For each Method 1, 2 \& 3, we have
\begin{equation}
\text{quality of solution} = \frac{\text{total distance without batching} - \text{objective value}}{\text{total distance without batching}}
\end{equation}
where the total distance without batching is computed with only the routing heuristic (which is the trivial batching objective value, and the best routing heuristic was used - Optimal solver), and the objective value is the heuristic solution of the JOBPRP obtained by the respective Methods 1, 2 \& 3. 
\end{definition}

\begin{remark}
A positive (resp. negative) quality of solution value indicates that there is a reduction (resp. increase) in objective value when a batching heuristic is used.
\end{remark}

\begin{example}
Using the same warehouse instance generated for Example \ref{tsh_example}, we have the optimal distance traveled (in meters) by each picker for each order $o$ (without batching) to be of the form $d_o^{(opt)}$, $o=1,2,\dots,7$, where $d_1^{(opt)} = 81, d_2^{(opt)} =87, d_3^{(opt)} =81, d_4^{(opt)} =78, d_5^{(opt)} =73, d_6^{(opt)} =88, d_7^{(opt)} =9$. Thus
\begin{equation*}
\text{total distance without batching} = \sum\limits_{o=1}^7 d_o^{(opt)} = 497
\end{equation*}
Recall from Example \ref{tsh_example} that the result of batching orders with TSH and S-shape as the router is $\{ \{3,5\}, \{4,6\}, \{2,7\}, \{1\} \}$. Then
\begin{equation*}
\text{objective value} = d_{35} + d_{46} + d_{27} + d_1 = 108 + 108 + 108 + 89 = 413
\end{equation*}
and consequently $\text{quality of solution} = \frac{497 - 413}{497} = 0.1690$.
{\flushright \qed}
\end{example}
\newpage

\section{Results For Each Warehouse Instance \& Method}

\subsection{Exact Solving}
Initially we submitted the job script to NUS HPC to solve the JOBPRP with constraints $(1)-(21)$ for large warehouse instances using the CBC solver, for all order instances in increments of $5$ (i.e. orders of sizes $5,10,\dots , 50$) with picker capacity $320$, and number of pickers available to be $40$ (as the highest number of pickers required from Methods $1$ - $3$ is $27$). We removed the time allowed for the solver to run, and noted that the runtime of each job (for solving JOBPRP to optimality for each order instance) was limited to $720$ hours, which is the wall-time of the job in HPC's parallel12 queue. We utilized $1$ unit of cpu, $32$GB memory for this job.

However every job, even after $240$ hours, did not attain even a suboptimal solution. This is beyond the practicalities of actual usage in the industry as in actual practice, it should not take too long for orders to be batched and picked as it will delay the delivery process to the customer. 

We decided to limit the solver runtime to $1$ hour for small warehouse instances and set the number of available pickers to the number computed from Method 2 with S-shape. This results in a lower upper bound in the number of pickers required, and reduces the search space in exact solving. In this case, some solutions were obtained, but for larger order instances, the runtime went beyond an hour so no solution was obtained. Thus, for the large warehouse instances existing results for complete optimal solving in \cite{valle2017} can be used for comparison instead, and we only show exact solving results for small warehouse instances with sufficient data points for analysis and focus on analyzing the results for heuristic solving with Methods 1 to 3.

\begin{figure}[h]
\includegraphics[width=0.49\textwidth]{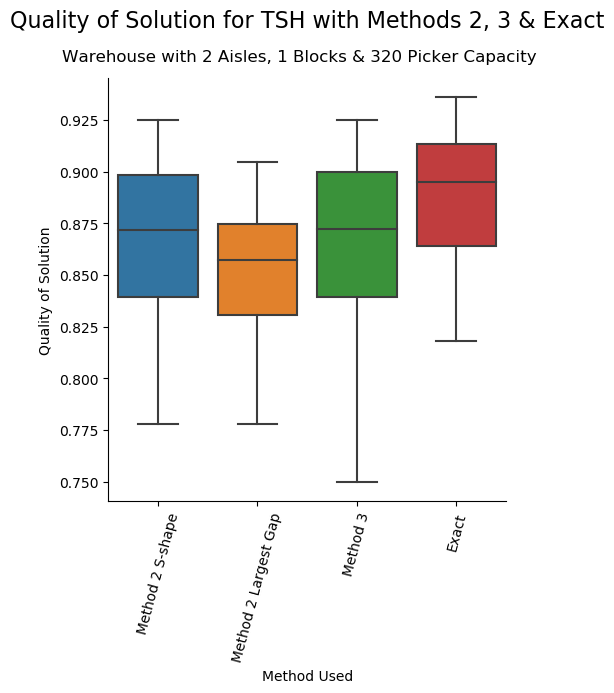}
\includegraphics[width=0.49\textwidth]{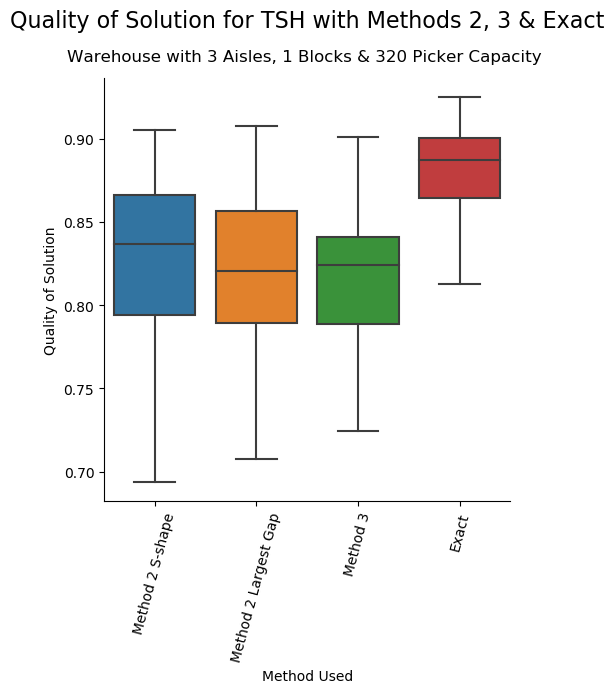}
\caption{Exact solving results for small warehouse instances with 2 \& 3 aisles}
\label{exact_solving_quality}
\end{figure}
\FloatBarrier

For exact solving, we were only able to obtain results with sufficient data points for small warehouse instances with 2 and 3 aisles. From the plots of their quality of solution, we observed that as the number of aisles increases, the gap between the median quality of solution for exact solving and heuristic solving with Methods 2 and 3 increases. No other generalizations can be made from the results obtained at the moment.

\subsection{Heuristic Solving (Large Instances)}

From the results obtained from heuristic solving with Methods 1 to 3, we made some general observations across all large warehouse instances, the first being that Method 1 Nearest Neighbor is the fastest (with lowest median time) across all blocks. This is not highly unusual, as in our implementation of the Nearest Neighbor router, there was no sequence of nodes (\code{node\_sequence}) to visit which contained several other nodes that are not pick nodes (like in S-shape and Largest Gap routers). This reduced the need for extensive computation of shortest paths, hence resulting in a significantly lower time to compute routes. The Nearest Neighbor router in Method 1 also resulted in the best (lowest) median objective value across all the other routers used in Method 1, except for 3 out of 16 results for Method 1. Note that the Nearest Neighbor heuristic is not used in practice in general, as actual aisles can be very narrow and it does not make sense for the picker to perform a u-turn in that narrow aisle, especially when the picker has a cart with them to store picks.

Another $\text{important}$ observation is that across all large warehouse instances and across all blocks, Method 2 (with its resp. routers) has a lower median objective value than Method 1 (with its resp. routers). This result is in line with the fact that using an optimal router instead of a routing heuristic for routing of pickers should give a objective value less than or equal to that of the routing heuristic. Thus, we decided to analyze only results from Methods without the Nearest Neighbor heuristic as a router, and excluded Method 1 from our analysis as its results were poorer than Methods 2 and 3.

To gain deeper insights on the results obtained for the total distance traveled by the pickers, we plotted the box plots of quality of solution against the method of solving used for each large warehouse instance and for 320 picker capacity. The box plots are of the form:
\begin{figure}[h]
\centering
\includegraphics[scale=.5]{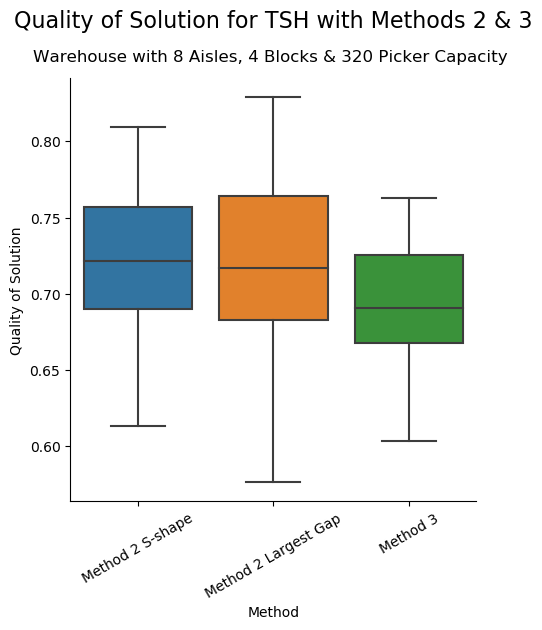}
\caption{Boxplot for Methods 2 and 3 of solving for quality of solution, for the warehouse instance with 8 aisles \& 4 blocks}
\label{all_boxplots_8_4}
\end{figure}
\FloatBarrier
One interesting yet perplexing observation we made was that there are methods with a higher median quality of solution and lower spread than Method 3. An example can be found in Figure~\ref{all_boxplots_8_4} above, where Method 3 has a lower median quality of solution than Method 2 with S-shape and Largest Gap router. This result is counter-intuitive as a natural thought will be that Method 3, having both savings and batch router to be optimal routers, should yield a higher quality of solution than all methods that employ routing heuristics in solving. Our interpretation of this result is that using a routing heuristic instead of an optimal router as savings router can result in batching being performed in such a way that the overall objective value will be lower than in the case where an optimal router is used as savings router. 

To gain a deeper insight as to what happened during the batching for Methods 2 and 3, we plotted the batches and routes for the test instance with $\Delta = 5, O = 10$ which had Method 2 with a higher quality of solution than Method 3. The following is an example of one such test instance with 10 orders where Method 2 with S-shape has a higher quality of solution than Method 3.

\begin{figure}[h]
\begin{minipage}{\linewidth}
\centering
\includegraphics[width=0.5\textwidth]{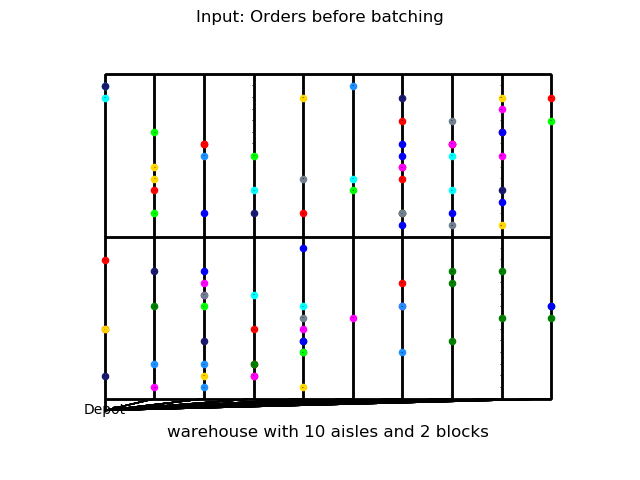}
\caption{Pick locations for each order (by color) before batching}
\end{minipage}
\vfill
\begin{minipage}{0.49\textwidth}
\centering
\includegraphics[width=0.98\textwidth]{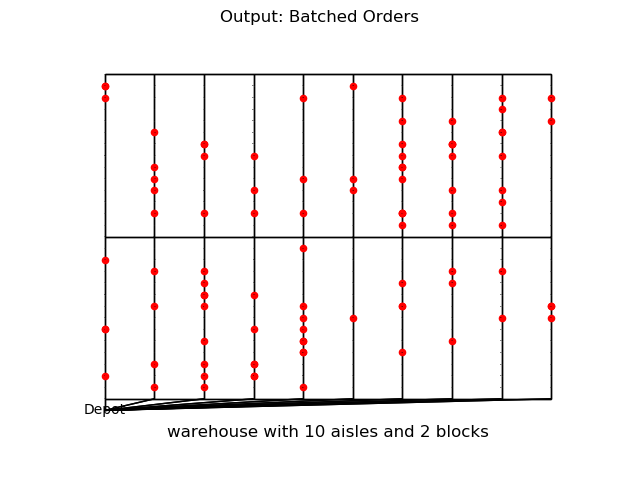}
\caption{Pick locations for each order (by color) after batching for Method 2}
\end{minipage}\hfill
\begin{minipage}{0.49\textwidth}
\centering
\includegraphics[width=0.98\textwidth]{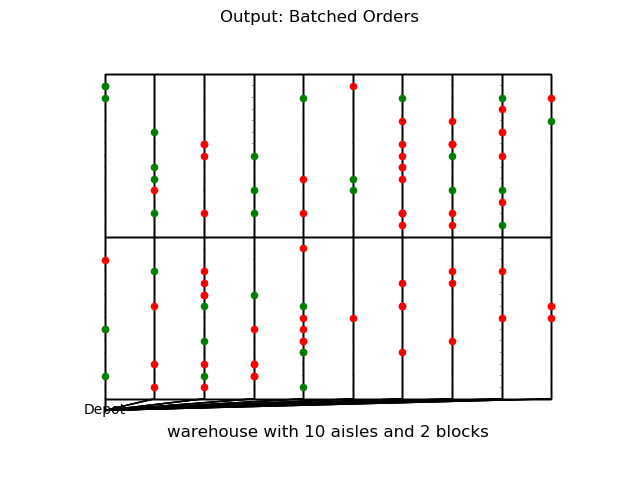}
\caption{Pick locations for each order (by color) after batching for Method 3}
\end{minipage}
\end{figure}

\begin{figure}[!htbp]
\begin{minipage}{\linewidth}
\centering
\includegraphics[width=0.5\textwidth]{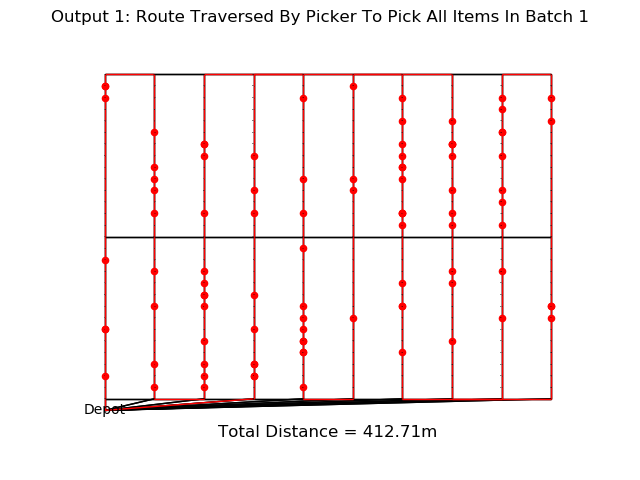}
\caption{Method 2's batched order route}
\label{m2_routed_batched_order}
\end{minipage}
\vfill
\begin{minipage}{0.49\textwidth}
\centering
\includegraphics[width=0.98\textwidth]{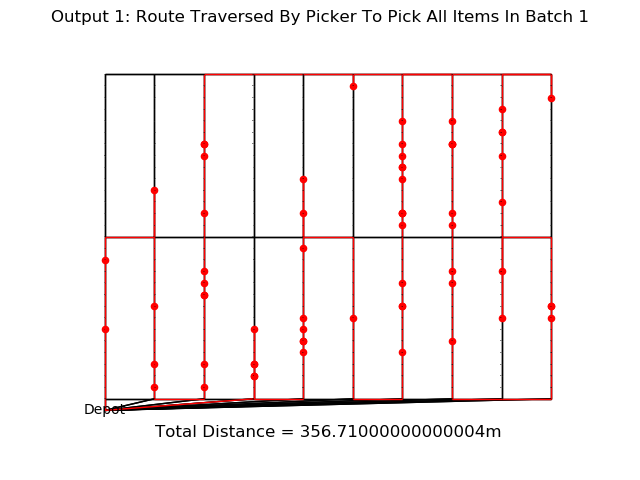}
\caption{Method 3's first batched order route}
\label{m3_first_routed_batched_order}
\end{minipage}\hfill
\begin{minipage}{0.49\textwidth}
\centering
\includegraphics[width=0.98\textwidth]{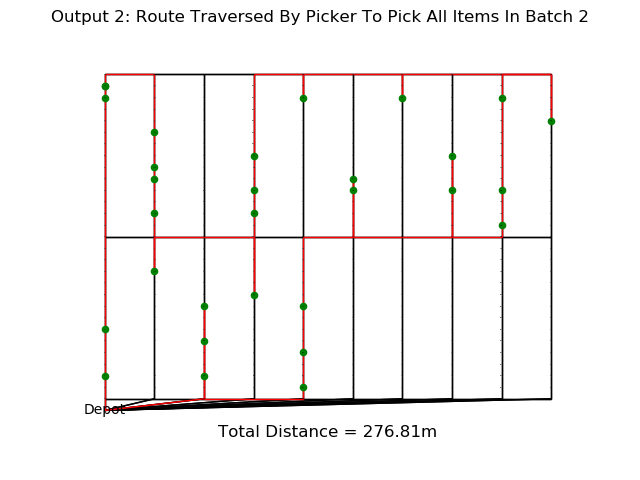}
\caption{Method 3's second batched order route}
\label{m3_second_routed_batched_order}
\end{minipage}
\end{figure}

In this test instance with $\Delta = 5, O = 10$, we observed that for an input with $10$ orders, Method 2 S-shape produced $1$ batched order as output whereas Method 3 produced $2$ batched orders as output. Furthermore, when these batched orders have been routed with the respective batch routers, the total distance traveled by the pickers across all the routed batched orders in Method 3 is more than that of Method 2 S-shape. The following plots in Figures \ref{m2_routed_batched_order}, \ref{m3_first_routed_batched_order} and \ref{m3_second_routed_batched_order} illustrates this.

\newpage

At this stage, we are not able to draw anything conclusive, so we decided to look into the critical point in the batching process where a new batch was formed for Method 3. Denote the orders by their indices from $1$ to $10$. Then we obtained the batched orders - for Method 2 S-shape: $\text{Batches} = \{\{3, 4, 1, 5, 9, 10, 2, 6, 8, 7\}\}$ and Method 3: $\text{Batches} = \{\{3, 4, 1, 2, 8, 7\}, \{5, 10, 6, 9\}\}$. Next, we computed the savings matrix for Method 2 S-shape and Method 3:

\newpage

\begin{table}[!htbp]
\resizebox{\textwidth}{!}{
\begin{tabular}{c|cccccccccc}
\multicolumn{1}{l|}{Order $i$} & 1 & 2 & 3 & 4 & 5 & 6 & 7 & 8 & 9 & 10 \\ \hline
1 & - &  &  &  &  &  &  &  &  &  \\
2 & 146.69 & - &  &  &  &  &  &  &  &  \\
3 & 156.47 & 156.26 & - &  &  &  &  &  &  &  \\
4 & 174.69 & 150.69 & 192.26 & - &  &  &  &  &  &  \\
5 & 173.09 & 130.69 & 152.47 & 154.69 & - &  &  &  &  &  \\
6 & 151.09 & 112.31 & 106.37 & 142.31 & 124.99 & - &  &  &  &  \\
7 & 100.9 & 86.69 & 100.47 & 100.69 & 86.9 & 72.52 & - &  &  &  \\
8 & 124.47 & 108.41 & 136.47 & 148.41 & 124.47 & 112.37 & 68.47 & - &  &  \\
9 & 147.09 & 128.69 & 140.47 & 144.69 & 165.09 & 124.71 & 92.9 & 116.47 & - &  \\
10 & 140.9 & 124.69 & 136.47 & 132.69 & 156.9 & 140.8 & 96.9 & 100.47 & 132.9 & -
\end{tabular}}
\caption{Method 2 S-shape Savings Matrix}
\label{m2_savings_matrix}
\end{table}

\begin{table}[!htbp]
\resizebox{\textwidth}{!}{
\begin{tabular}{c|cccccccccc}
Order $i$ & 1 & 2 & 3 & 4 & 5 & 6 & 7 & 8 & 9 & 10 \\ \hline
1 & - &  &  &  &  &  &  &  &  &  \\
2 & 96.8 & - &  &  &  &  &  &  &  &  \\
3 & 128.47 & 130.47 & - &  &  &  &  &  &  &  \\
4 & 142.42 & 112.8 & 144.37 & - &  &  &  &  &  &  \\
5 & 126.71 & 94.52 & 110.47 & 114.52 & - &  &  &  &  &  \\
6 & 112.71 & 92.71 & 92.52 & 104.42 & 100.71 & - &  &  &  &  \\
7 & 100.31 & 78.59 & 88.41 & 108.59 & 100.41 & 84.31 & - &  &  &  \\
8 & 104.47 & 92.47 & 118.47 & 126.47 & 100.47 & 104.37 & 88.41 & - &  &  \\
9 & 124.71 & 90.52 & 112.47 & 112.52 & 124.81 & 120.43 & 92.69 & 90.19 & - &  \\
10 & 126.47 & 110.37 & 122.47 & 120.47 & 132.47 & 128.71 & 92.26 & 112.47 & 124.19 & -
\end{tabular}}
\caption{Method 3 Savings Matrix}
\label{m3_savings_matrix}
\end{table}

So what happened during the batching? To find out more, we computed the steps of the batching, where we found that order 5 does not get added to the first batch of orders at the third step:

\begin{center}
\begin{minipage}{2.1in}
\begin{tabular}{l}
$\textbf{Step 1}: \{\{3,4\}\}$ \\
$\textbf{Step 2}: \{\{3,4,1\}\}$ \\
$\textbf{Step 3}: \{\{3,4,1,5\}\}$
\end{tabular}
\begin{center}
Steps of Method 2's batching
\end{center}
\end{minipage}
\quad
\begin{minipage}{2.1in}
\begin{tabular}{l}
$\textbf{Step 1}: \{\{3,4\}\}$ \\
$\textbf{Step 2}: \{\{3,4,1\}\}$ \\
$\textbf{Step 3}: \{\{3,4,1\}, \{5,10\}\}$
\end{tabular}
\begin{center}
Steps of Method 3's batching
\end{center}
\end{minipage}
\end{center}

\newpage

\noindent To analyze this critical step deeper, we refer to the savings matrix for Method 3:

\begin{table}[h]
\resizebox{\textwidth}{!}{
\begin{tabular}{c|cccccccccc}
Order $i$ & 1 & 2 & 3 & 4 & 5 & 6 & 7 & 8 & 9 & 10 \\ \hline
1 & - &  &  &  &  &  &  &  &  &  \\
2 & 96.8 & - &  &  &  &  &  &  &  &  \\
3 & 128.47 & 130.47 & - &  &  &  &  &  &  &  \\
4 & 142.42 & 112.8 & 144.37 & - &  &  &  &  &  &  \\
5 & \textcolor{blue}{126.71} & 94.52 & 110.47 & 114.52 & - &  &  &  &  &  \\
6 & 112.71 & 92.71 & 92.52 & 104.42 & 100.71 & - &  &  &  &  \\
7 & 100.31 & 78.59 & 88.41 & 108.59 & 100.41 & 84.31 & - &  &  &  \\
8 & 104.47 & 92.47 & 118.47 & 126.47 & 100.47 & 104.37 & 88.41 & - &  &  \\
9 & 124.71 & 90.52 & 112.47 & 112.52 & 124.81 & 120.43 & 92.69 & 90.19 & - &  \\
10 & 126.47 & 110.37 & 122.47 & 120.47 & \textcolor{red}{132.47} & 128.71 & 92.26 & 112.47 & 124.19 & -
\end{tabular}}
\caption{Method 3 Savings Matrix With Highlighted Key Savings}
\label{m3_savings_matrix_2}
\end{table}

In the third step for Method 3, we observed that order pair $(5,10)$ had a higher savings than $(1,5)$ in the third step, i.e. $s_{5,10} = \textcolor{red}{132.47} \geq \textcolor{blue}{126.71} = s_{1,5}$. On the contrary for Method 2 S-shape, it can be seen from Table \ref{m2_savings_matrix} that order pair $(5,10)$ had a lower savings than $(1,5)$. As a result, a new batch is formed for order pair $(5,10)$ in step 3 for Method 3, which ultimately lead to a higher total distance traveled for Method 3.

We also computed the median time elapsed for each method of solving, by combining all same methods of solving with the S-shape and Largest Gap routers together. For e.g., we analyzed the results of Method 1 with S-shape, Largest Gap for the savings router together. We obtained the following plot as a result:

\begin{figure}[h]
\centering
\includegraphics[scale=.7]{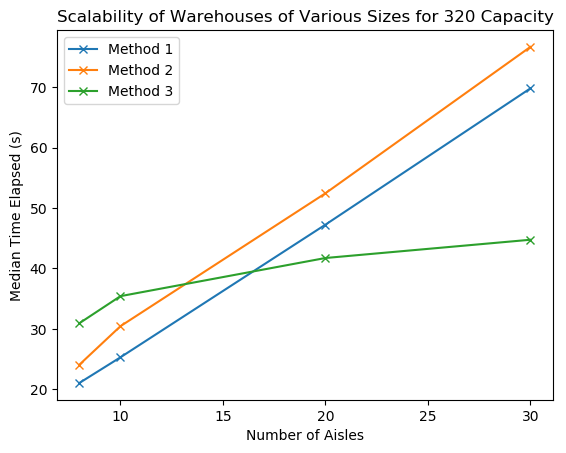}
\caption{Plot of Median Time Elapsed in seconds across all large warehouse instances for 320 picker capacity}
\label{warehouse_time_scalability}
\end{figure}
\FloatBarrier

Another important yet surprising finding in our experimental results is that the median time taken for Method 3's solving is the lowest for warehouses with 20 and 30 aisles as compared to 8 and 10 aisles. At this point in time, we are not able to explain why that is the case for Method 3, but it may be possible that the number of aisles also had a role in influencing the batching process (besides the savings router which differs across methods).

We also looked into how large an impact varying the number of blocks of each warehouse instance has on the median total distance traveled. Like with the case for median time elapsed, we analyzed each method of solving as a whole, across all the savings routers. 

\begin{figure}[h]
\centering
\includegraphics[scale=.7]{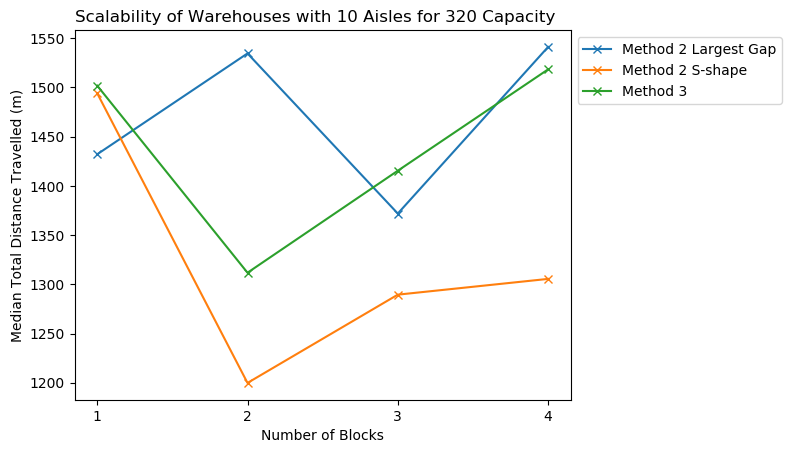}
\caption{Plot of Median Total Distance Travelled across all number of blocks for a warehouse with 10 aisles and 320 picker capacity}
\label{warehouse_objective_scalability}
\end{figure}
\FloatBarrier

From the results collected, we observed no clear result across Methods 2 and 3 when the number of blocks for each warehouse instance was varied. Increasing the number of blocks does not lead to a consistent increase (resp. decrease) in median total distance traveled. An interesting observation we made is that there are several warehouse instances having Method 2 (with either S-shape or Largest Gap router) produce a solution with a lower median total distance traveled than Method 3. An example can be found in Figure \ref{warehouse_objective_scalability} above, where Method 2 with S-shape router produced a lower median total distance traveled than Method 3 across all blocks for a warehouse with 10 aisles. 

Recommendations on the 'best' method of solving for each warehouse instance with varying number of blocks will be made in a subsequent section.

\newpage

\section{Recommendations}
In this section, we present our answers to the two main questions in the thesis. From our results, one clear result is that Method 1 (regardless of routing heuristic) had the worst median quality of solution. However, its median computation time was the smallest among all the Methods for warehouses with 1 block, with the same routing heuristic in Methods 1 and 2. Method 2 (with S-shape or Largest Gap) had the best quality of solution versus time trade-off across all warehouse instances. Method 3 also has a very good quality of solution versus time trade-off, as it has consistent results in median time elapsed and high quality of solution for most instances. Method 2 with S-shape and Largest Gap as savings routers on the other hand also show similar results for quality of solution like in Figure~\ref{all_boxplots_8_4} and also median time elapsed across most warehouse instances. However, all methods of solving involving S-shape and Largest Gap routers (batch and/or savings) have large computation times, which become relatively large as the number of aisles in the warehouse increases.

Finally, we come to the second and last question - If I have a warehouse with $X$ aisles and $Y$ cross-aisles and a number of orders to batch, what method should be used to solve the problem within the time limit? Reinforcing our findings for the quality of solution above with the tables in the appendix where methods with the lowest median objective values are highlighted in \textcolor{red}{red}, we have the following recommendations.

We first make a remark that the number of blocks in the warehouse is equal to $Y$ cross-aisles $- 1$. At this stage, if computation time is the primary concern (regardless of objective value), then the best method for all warehouse instances will be a warehouse with 1 block, and Method 1 with any routing heuristic (S-shape or Largest Gap) as their median computation time differs by at most 5 seconds across all warehouse instances. If objective value is the primary concern, then the recommended methods based on our results can be found in the table below:

\begin{longtable}[c]{|c|c|c|}
\hline
\textbf{Aisles} & \textbf{Blocks} & \textbf{Recommended Method} \\ \hline
\endfirsthead
\endhead
8               & 1               & Method 3   \\ \hline
8               & 2               & Method 3   \\ \hline
8               & 3               & Method 2 Largest Gap        \\ \hline
8               & 4               & Method 2 Largest Gap        \\ \hline
10              & 1               & Method 2 Largest Gap      \\ \hline
10              & 2               & Method 2 S-shape            \\ \hline
10              & 3               & Method 2 S-shape            \\ \hline
10              & 4               & Method 2 S-shape            \\ \hline
20              & 1               & Method 2 Largest Gap        \\ \hline
20              & 2               & Method 2 Largest Gap        \\ \hline
20              & 3               & Method 3                    \\ \hline
20              & 4               & Method 2 S-shape            \\ \hline
30              & 1               & Method 2 Largest Gap        \\ \hline
30              & 2               & Method 2 S-shape            \\ \hline
30              & 3               & Method 3                    \\ \hline
30              & 4               & Method 2 S-shape            \\ \hline
\end{longtable}

\section{Conclusion}
In this thesis, we investigated the heuristic methods of order batching and picker routing in a simulated warehouse environment to answer the two main questions in our business problem. We implemented a batching heuristic (TSH) and four routers - Nearest Neighbor, S-shape, Largest Gap and Optimal. We demonstrated the benefits of the different methods of heuristic solving for the JOBPRP, where we contrasted the trade-offs between quality of solution and computation time as an answer to the first question, and showed that Method 2 is predominantly the most recommended method as an answer to the second question. We also showed that Method 3 is not always the method of solving with the highest quality of solution, which is counter-intuitive. 

\section{Future Work}
Due to the time constraint of the project, we were only able to implement one batching heuristic to address order batching. For our next step, we will be looking into developing a new batching heuristic, which formulates batching as a covering problem (like a vertex cover), but with intervals which are subsets of the routes (i.e. first and last pick locations that lie on the pick path) containing the original order. A rough outline of the heuristic with a running example is as follows:

\begin{enumerate}
\item Given a list of orders, e.g. $O = \{\{2,3,7\}, \{4,4,6\}, \{9\}, \{5,7,8,8,8\}\}$, first combine all orders into one merged order: $O' = \{2,3,4,4,5,6,7,7,8,8,8,9\}$.
\item Perform optimal routing on $O'$ (with PyConcorde), which gives us a sorted sequence of items/locations.
\item Each original order can now be considered to span a certain interval along the routing of the merged order, i.e. represented by an interval $[i,j]$.
\item Batch intervals, starting with one with the greatest overlap. The intuition being that two order intervals overlap if items are being picked together in the merged order.
\end{enumerate}

We also noticed that the run-time for Methods 1 to 3 for S-shape and Largest Gap were still much longer than the time elapsed for similar methods employed in solving in~\cite{valle2017}. We aim to improve the run-time by first reducing the need for the shortest paths to all nodes to be computed in the \code{route} function in the routers. Currently, the shortest path and its distance for each picker is being computed by finding the shortest path between consecutive nodes in \code{node\_sequence}, which is a sequence of nodes to be traversed. Note that due to the nature of some routers, the nodes in \code{node\_sequence} may not be in sequence. As a next step, we will be looking into the routing heuristics to see where more information can be given such that more nodes can be appended to \code{sequence} in sequence. Once all nodes are in sequence, we can simply employ a function \code{get\_distance} which returns the distance between consecutive nodes, by extracting the edge weight from the graph \code{G} which was initially set up and already contains all the weighted edges in the warehouse. The function can be implemented as follows:

\begin{python}
def get_distance(route, G):
    return sum([G[s][t]['weight'] for s, t in zip(route, route[1:])])
\end{python}

\end{doublespacing}

\bibliographystyle{unsrt}\nocite{*}
\bibliography{references}

\begin{thebibliography}{10}

\bibitem{valle2017}
Cristiano~Arbex Valle, John~E. Beasley, and Alexandre~Salles da~Cunha.
\newblock Optimally solving the joint order batching and picker routing
  problem.
\newblock {\em European Journal of Operational Research}, 262(3):817 -- 834,
  2017.

\bibitem{roodbergen2001}
Kees~Jan Roodbergen and RenÉde Koster.
\newblock Routing methods for warehouses with multiple cross aisles.
\newblock {\em International Journal of Production Research}, 39(9):1865--1883,
  2001.

\bibitem{bartholdi2017}
John J~Bartholdi and Steven T~Hackman.
\newblock Warehouse \& distribution science release 0.98.
\newblock 08 2017.

\bibitem{koster2007}
René de~Koster, Tho Le-Duc, and Kees~Jan Roodbergen.
\newblock Design and control of warehouse order picking: A literature review.
\newblock {\em European Journal of Operational Research}, 182(2):481 -- 501,
  2007.

\bibitem{gademann2005}
NOUD GADEMANN and VAN DE~STEEF VELDE.
\newblock Order batching to minimize total travel time in a parallel-aisle
  warehouse.
\newblock {\em IIE Transactions}, 37(1):63--75, 2005.

\bibitem{scholz2016}
André Scholz, Sebastian Henn, Meike Stuhlmann, and Gerhard Wäscher.
\newblock A new mathematical programming formulation for the single-picker
  routing problem.
\newblock {\em European Journal of Operational Research}, 253(1):68 -- 84,
  2016.

\bibitem{wolsey1998}
Wolsey~L. A.
\newblock Integer programming.
\newblock 1998.

\bibitem{rothlauf2011}
Franz Rothlauf.
\newblock {\em Design of Modern Heuristics: Principles and Application}.
\newblock Springer Publishing Company, Incorporated, 1st edition, 2011.

\bibitem{applegate2007}
David~L. Applegate, Robert~E. Bixby, Vasek Chvatal, and William~J. Cook.
\newblock {\em The Traveling Salesman Problem: A Computational Study (Princeton
  Series in Applied Mathematics)}.
\newblock Princeton University Press, Princeton, NJ, USA, 2007.

\bibitem{hwang2005}
D.~G.~Kim H.~Hwang.
\newblock Order-batching heuristics based on cluster analysis in a low-level
  picker-to-part warehousing system.
\newblock {\em International Journal of Production Research},
  43(17):3657--3670, 2005.

\bibitem{ho2008}
Ying-Chin Ho, Teng-Sheng Su, and Zhi-Bin Shi.
\newblock Order-batching methods for an order-picking warehouse with two cross
  aisles.
\newblock {\em Computers \& Industrial Engineering}, 55(2):321 -- 347, 2008.

\bibitem{koster1999}
René De~Koster, Poort E.S.V.D, and Wolters M.
\newblock Efficient orderbatching methods in warehouses.
\newblock 37:1479--1504, 05 1999.

\bibitem{clarke1964}
G.~Clarke and J.~W. Wright.
\newblock Scheduling of vehicles from a central depot to a number of delivery
  points.
\newblock {\em Operations Research}, 12(4):568--581, 1964.

\bibitem{elsayed1989}
E.~A. ELSAYED and O.~I. UNAL.
\newblock Order batching algorithms and travel-time estimation for automated
  storage/retrieval systems.
\newblock {\em International Journal of Production Research}, 27(7):1097--1114,
  1989.

\bibitem{roodbergen2008}
Gunter P.~Sharp Kees Jan~Roodbergen and Iris~F.A. Vis.
\newblock Designing the layout structure of manual order picking areas in
  warehouses.
\newblock {\em IIE Transactions}, 40(11):1032--1045, 2008.

\end{thebibliography}

\appendix

\chapter{Appendix}

\begin{doublespacing}

\section{Heuristic Solving Results For Large Instances}
The tables in this section are the results of Methods 1 to 3 showing the median time taken to get the solution, and also the median objective value, for different routers (S-shape, Largest Gap, Optimal). We highlighted the smallest median time and objective value amongst the methods for each warehouse instance in \textcolor{red}{red}. Results for Methods which used Nearest Neighbor as the router are not shown as the heuristic is not used in practice due to practical reasons described in Chapter 5.

\end{doublespacing}

\begin{table}[!htb]
\centering
\caption{Results for warehouses with 8 aisles, 1 to 4 blocks.}\label{8aisles1to4blocks}
\scriptsize
\begin{subtable}{1.0\textwidth}
\sisetup{table-format=-1.2}   
\centering
\begin{tabular}{|c|c|c|}
\hline
 & \textbf{Median Time} & \textbf{Median Objective Value} \\ \hline
\textbf{Method 1 S-shape} & \textcolor{red}{11.35} & 1413.36 \\ \hline
\textbf{Method 1 L-Gap} & 11.57 & 1828.92 \\ \hline
\textbf{Method 2 S-shape} & 12.85 & 1393.38 \\ \hline
\textbf{Method 2 L-Gap} & 13.14 & 1428.79 \\ \hline
\textbf{Method 3} & 34.2 & \textcolor{red}{1392.84} \\ \hline
\end{tabular}
\caption{1 Block}\label{8aisle1block}
\end{subtable}

\bigskip
\begin{subtable}{1.0\textwidth}
\sisetup{table-format=4.0} 
\centering
\begin{tabular}{|c|c|c|}
\hline
 & \textbf{Median Time} & \textbf{Median Objective Value} \\ \hline
\textbf{Method 1 S-shape} & 20.19 & 1541.17 \\ \hline
\textbf{Method 1 L-Gap} & \textcolor{red}{19.84} & 1853.62 \\ \hline
\textbf{Method 2 S-shape} & 22.32 & 1442.41 \\ \hline
\textbf{Method 2 L-Gap} & 23.16 & 1382.62 \\ \hline
\textbf{Method 3} & 31.75 & \textcolor{red}{1374.62} \\ \hline
\end{tabular}
\caption{2 Blocks}\label{8aisle2blocks}
\end{subtable}

\bigskip
\begin{subtable}{1.0\textwidth}
\sisetup{table-format=4.0} 
\centering
\begin{tabular}{|c|c|c|}
\hline
\textbf{} & \textbf{Median Time} & \textbf{Median Objective Value} \\ \hline
\textbf{Method 1 S-shape} & 29.69 & 1622.77 \\ \hline
\textbf{Method 1 L-Gap} & \textcolor{red}{28.34} & 1685.52 \\ \hline
\textbf{Method 2 S-shape} & 34.99 & 1343.69 \\ \hline
\textbf{Method 2 L-Gap} & 35.69 & \textcolor{red}{1235.17} \\ \hline
\textbf{Method 3} & 29.77 & 1309.26 \\ \hline
\end{tabular}
\caption{3 Blocks}\label{8aisles3blocks}
\end{subtable}

\bigskip
\begin{subtable}{1.0\textwidth}
\sisetup{table-format=4.0} 
\centering
\begin{tabular}{|c|c|c|}
\hline
 & \textbf{Median Time} & \textbf{Median Objective Value} \\ \hline
\textbf{Method 1 S-shape} & 38.37 & 1584.11 \\ \hline
\textbf{Method 1 L-Gap} & 36.7 & 1810.31 \\ \hline
\textbf{Method 2 S-shape} & 40.6 & 1307.17 \\ \hline
\textbf{Method 2 L-Gap} & 42.0 & \textcolor{red}{1235.08} \\ \hline
\textbf{Method 3} & \textcolor{red}{29.55} & 1488.98 \\ \hline
\end{tabular}
\caption{4 Blocks}\label{8aisles4blocks}
\end{subtable}

\end{table}
\FloatBarrier

\begin{table}[!htb]
\centering
\caption{Results for warehouses with 10 aisles, 1 to 4 blocks.}\label{10aisles1to4blocks}
\scriptsize
\begin{subtable}{1.0\textwidth}
\sisetup{table-format=-1.2}   
\centering
\begin{tabular}{|c|c|c|}
\hline
\textbf{} & \textbf{Median Time} & \textbf{Median Objective Value} \\ \hline
\textbf{Method 1 S-shape} & \textcolor{red}{13.1} & 1530.28 \\ \hline
\textbf{Method 1 L-Gap} & 13.71 & 1762.89 \\ \hline
\textbf{Method 2 S-shape} & 16.96 & 1494.08 \\ \hline
\textbf{Method 2 L-Gap} & 15.85 & \textcolor{red}{1432.08} \\ \hline
\textbf{Method 3} & 44.14 & 1501.82 \\ \hline
\end{tabular}
\caption{1 Block}\label{10aisle1block}
\end{subtable}

\bigskip
\begin{subtable}{1.0\textwidth}
\sisetup{table-format=4.0} 
\centering
\begin{tabular}{|c|c|c|}
\hline
 & \textbf{Median Time} & \textbf{Median Objective Value} \\ \hline
\textbf{Method 1 S-shape} & 24.54 & 1231.27 \\ \hline
\textbf{Method 1 L-Gap} & \textcolor{red}{24.11} & 1992.94 \\ \hline
\textbf{Method 2 S-shape} & 29.93 & \textcolor{red}{1199.85} \\ \hline
\textbf{Method 2 L-Gap} & 28.59 & 1534.53 \\ \hline
\textbf{Method 3} & 39.27 & 1311.79 \\ \hline
\end{tabular}
\caption{2 Blocks}\label{10aisle2blocks}
\end{subtable}

\bigskip
\begin{subtable}{1.0\textwidth}
\sisetup{table-format=4.0} 
\centering
\begin{tabular}{|c|c|c|}
\hline
\textbf{} & \textbf{Median Time} & \textbf{Median Objective Value} \\ \hline
\textbf{Method 1 S-shape} & 35.9 & 1646.7 \\ \hline
\textbf{Method 1 L-Gap} & 33.59 & 1975.24 \\ \hline
\textbf{Method 2 S-shape} & 37.89 & \textcolor{red}{1289.51} \\ \hline
\textbf{Method 2 L-Gap} & 46.67 & 1371.89 \\ \hline
\textbf{Method 3} & \textcolor{red}{31.2} & 1415.39 \\ \hline
\end{tabular}
\caption{3 Blocks}\label{10aisles3blocks}
\end{subtable}

\bigskip
\begin{subtable}{1.0\textwidth}
\sisetup{table-format=4.0} 
\centering
\begin{tabular}{|c|c|c|}
\hline
 & \textbf{Median Time} & \textbf{Median Objective Value} \\ \hline
\textbf{Method 1 S-shape} & 46.31 & 1582.81 \\ \hline
\textbf{Method 1 L-Gap} & 45.43 & 2319.65 \\ \hline
\textbf{Method 2 S-shape} & 51.38 & \textcolor{red}{1305.51} \\ \hline
\textbf{Method 2 L-Gap} & 48.97 & 1541.12 \\ \hline
\textbf{Method 3} & \textcolor{red}{30.46} & 1518.32 \\ \hline
\end{tabular}
\caption{4 Blocks}\label{10aisles4blocks}
\end{subtable}

\end{table}
\FloatBarrier

\begin{table}[!htb]
\centering
\caption{Results for warehouses with 20 aisles, 1 to 4 blocks.}\label{20aisles1to4blocks}
\scriptsize
\begin{subtable}{1.0\textwidth}
\sisetup{table-format=-1.2}   
\centering
\begin{tabular}{|c|c|c|}
\hline
 & \textbf{Median Time} & \textbf{Median Objective Value} \\ \hline
\textbf{Method 1 S-shape} & \textcolor{red}{23.65} & 1798.22 \\ \hline
\textbf{Method 1 L-Gap} & 24.02 & 1836.94 \\ \hline
\textbf{Method 2 S-shape} & 28.87 & 1715.45 \\ \hline
\textbf{Method 2 L-Gap} & 31.86 & \textcolor{red}{1657.33} \\ \hline
\textbf{Method 3} & 44.18 & 1828.91 \\ \hline
\end{tabular}
\caption{1 Block}\label{20aisle1block}
\end{subtable}

\bigskip
\begin{subtable}{1.0\textwidth}
\sisetup{table-format=4.0} 
\centering
\begin{tabular}{|c|c|c|}
\hline
 & \textbf{Median Time} & \textbf{Median Objective Value} \\ \hline
\textbf{Method 1 S-shape} & 45.86 & 1965.17 \\ \hline
\textbf{Method 1 L-Gap} & 45.24 & 2326.09 \\ \hline
\textbf{Method 2 S-shape} & 53.67 & 1819.23 \\ \hline
\textbf{Method 2 L-Gap} & 48.94 & \textcolor{red}{1619.94} \\ \hline
\textbf{Method 3} & \textcolor{red}{43.28} & 1690.73 \\ \hline
\end{tabular}
\caption{2 Blocks}\label{20aisle2blocks}
\end{subtable}

\bigskip
\begin{subtable}{1.0\textwidth}
\sisetup{table-format=4.0} 
\centering
\begin{tabular}{|c|c|c|}
\hline
 & \textbf{Median Time} & \textbf{Median Objective Value} \\ \hline
\textbf{Method 1 S-shape} & 68.97 & 2565.05 \\ \hline
\textbf{Method 1 L-Gap} & 67.17 & 2900.35 \\ \hline
\textbf{Method 2 S-shape} & 72.79 & 1838.04 \\ \hline
\textbf{Method 2 L-Gap} & 69.7 & 1755.7 \\ \hline
\textbf{Method 3} & \textcolor{red}{39.79} & \textcolor{red}{1572.88} \\ \hline
\end{tabular}
\caption{3 Blocks}\label{20aisles3blocks}
\end{subtable}

\bigskip
\begin{subtable}{1.0\textwidth}
\sisetup{table-format=4.0} 
\centering
\begin{tabular}{|c|c|c|}
\hline
 & \textbf{Median Time} & \textbf{Median Objective Value} \\ \hline
\textbf{Method 1 S-shape} & 88.5 & 2069.98 \\ \hline
\textbf{Method 1 L-Gap} & 87.43 & 3380.06 \\ \hline
\textbf{Method 2 S-shape} & 95.5 & \textcolor{red}{1494.04} \\ \hline
\textbf{Method 2 L-Gap} & 91.82 & 1807.75 \\ \hline
\textbf{Method 3} & \textcolor{red}{38.43} & 1767.76 \\ \hline
\end{tabular}
\caption{4 Blocks}\label{20aisles4blocks}
\end{subtable}

\end{table}
\FloatBarrier

\begin{table}[!htb]
\centering
\caption{Results for warehouses with 30 aisles, 1 to 4 blocks.}\label{30aisles1to4blocks}
\scriptsize
\begin{subtable}{1.0\textwidth}
\sisetup{table-format=-1.2}   
\centering
\begin{tabular}{|c|c|c|}
\hline
 & \textbf{Median Time} & \textbf{Median Objective Value} \\ \hline
\textbf{Method 1 S-shape} & 35.65 & 2202.18 \\ \hline
\textbf{Method 1 L-Gap} & \textcolor{red}{35.08} & 1925.62 \\ \hline
\textbf{Method 2 S-shape} & 38.55 & 2073.26 \\ \hline
\textbf{Method 2 L-Gap} & 43.29 & \textcolor{red}{1803.82} \\ \hline
\textbf{Method 3} & 42.03 & 1950.79 \\ \hline
\end{tabular}
\caption{1 Block}\label{30aisle1block}
\end{subtable}

\bigskip
\begin{subtable}{1.0\textwidth}
\sisetup{table-format=4.0} 
\centering
\begin{tabular}{|c|c|c|}
\hline
 & \textbf{Median Time} & \textbf{Median Objective Value} \\ \hline
\textbf{Method 1 S-shape} & 67.15 & 1939.31 \\ \hline
\textbf{Method 1 L-Gap} & 67.04 & 2779.71 \\ \hline
\textbf{Method 2 S-shape} & 72.92 & \textcolor{red}{1797.22} \\ \hline
\textbf{Method 2 L-Gap} & 73.43 & 1953.58 \\ \hline
\textbf{Method 3} & \textcolor{red}{47.83} & 2129.55 \\ \hline
\end{tabular}
\caption{2 Blocks}\label{30aisle2blocks}
\end{subtable}

\bigskip
\begin{subtable}{1.0\textwidth}
\sisetup{table-format=4.0} 
\centering
\begin{tabular}{|c|c|c|}
\hline
 & \textbf{Median Time} & \textbf{Median Objective Value} \\ \hline
\textbf{Method 1 S-shape} & 101.64 & 2994.09 \\ \hline
\textbf{Method 1 L-Gap} & 98.78 & 3560.48 \\ \hline
\textbf{Method 2 S-shape} & 104.52 & 2061.69 \\ \hline
\textbf{Method 2 L-Gap} & 105.72 & 2021.51 \\ \hline
\textbf{Method 3} & \textcolor{red}{44.88} & \textcolor{red}{1932.47} \\ \hline
\end{tabular}
\caption{3 Blocks}\label{30aisles3blocks}
\end{subtable}

\bigskip
\begin{subtable}{1.0\textwidth}
\sisetup{table-format=4.0} 
\centering
\begin{tabular}{|c|c|c|}
\hline
 & \textbf{Median Time} & \textbf{Median Objective Value} \\ \hline
\textbf{Method 1 S-shape} & 133.35 & 2741.11 \\ \hline
\textbf{Method 1 L-Gap} & 128.82 & 4338.02 \\ \hline
\textbf{Method 2 S-shape} & 133.01 & \textcolor{red}{1937.99} \\ \hline
\textbf{Method 2 L-Gap} & 135.99 & 2071.48 \\ \hline
\textbf{Method 3} & \textcolor{red}{45.29} & 2037.7 \\ \hline
\end{tabular}
\caption{4 Blocks}\label{30aisles4blocks}
\end{subtable}

\end{table}
\FloatBarrier

\end{document}